\documentclass{emulateapj}
\usepackage{graphicx}
\usepackage[space]{grffile}
\usepackage{latexsym}
\usepackage{textcomp}
\usepackage{longtable}
\usepackage{multirow,booktabs}
\usepackage{amsfonts,amsmath,amssymb}
\usepackage{natbib}
\usepackage{url}
\usepackage{hyperref}
\hypersetup{colorlinks=false,pdfborder={0 0 0}}

\usepackage[english]{babel}

\usepackage[utf8]{inputenc}

\newcommand{\mso}{\,\mathrm{M}_\odot}

\shorttitle{
Asteroseismic signatures of evolving internal stellar magnetic fields}
\shortauthors{Matteo Cantiello}

\begin{document}


\title{
Asteroseismic signatures of evolving internal stellar magnetic fields}


\author{%
Matteo Cantiello,\altaffilmark{\dag,1} 
Jim Fuller,\altaffilmark{\dag,1,2} 
Lars Bildsten\altaffilmark{1,3}
}
\altaffiltext{\dag}{These authors contributed equally to this work}
\altaffiltext{1}{Kavli Institute for Theoretical Physics, University of California, Santa Barbara, CA 93106, USA}
\altaffiltext{2}{TAPIR, Walter Burke Institute for Theoretical Physics, Mailcode 350-17, California Institute of Technology, Pasadena, CA 91125, USA}
\altaffiltext{3}{Department of Physics, University of California, Santa Barbara, CA 93106, USA}

\email{Corresponding author: matteo@kitp.ucsb.edu}




\begin{abstract}
Recent asteroseismic analyses have revealed the presence of strong (B $\gtrsim 10^5$ G) magnetic fields in the cores of many red giant stars. 
Here, we examine the implications of these results for the  evolution of stellar  magnetic fields, and we make predictions for future observations. Those stars with suppressed dipole modes indicative of strong core fields should exhibit moderate but detectable quadrupole mode suppression.
The long magnetic diffusion times within stellar cores ensure that dynamo-generated fields are confined to mass coordinates within the main sequence convective core, and the observed sharp increase in dipole mode suppression rates above $1.5 \, M_\odot$ may be explained by the larger convective core masses and faster rotation of these more massive stars. In clump stars, core fields of $\sim10^5 \, {\rm G}$ can suppress dipole modes, whose visibility should be equal to or less than the visibility of suppressed modes in ascending red giants. High dipole mode suppression rates in low-mass ($M \lesssim 2 \, M_\odot$) clump stars would indicate that magnetic fields generated during the main sequence can withstand subsequent convective phases and survive into the compact remnant phase. Finally, we discuss implications for observed magnetic fields in white dwarfs and neutron stars, as well as the effects of magnetic fields in various types of pulsating stars.
  
\end{abstract}%


\bibliographystyle{apj}

\section{Introduction}\label{intro}
Magnetic fields are ubiquitous in astrophysics. From  galaxy clusters to strongly magnetized neutron stars (magnetars) their amplitude spans more than 20 orders of magnitudes \citep{Brandenburg_2005}. Using the Zeeman effect, stellar magnetic fields have been measured at the surface of stars
for more than 60 years \citep{Babcock_1947,Landstreet_1992,Donati_2009}. Stars with convective envelopes show the presence of surface magnetic fields that are believed to be produced by contemporary dynamo action. A small fraction of stars with radiative envelopes also have strong ($B \! \sim \! {\rm kG}$) large scale magnetic fields that are likely generated or inherited during the star formation process (fossil fields). This fraction appears to be 5-10\% for main sequence (MS) A and OB stars \citep[e.g.,][]{Auriere2004,2012ASPC..464..405W}.
On the other hand, weak ($B \! \lesssim \! 100 \, {\rm G}$) and/or small scale fields could be much more common at the surface of A and OB stars \citep{Cantiello_2009,Cantiello_2011,Braithwaite_2012}. 
Until recently, evidence for stellar magnetism has been limited to surface fields, with only compact remnants providing clues about the level of internal magnetization of their progenitor stars.

Thanks to a new asteroseismic technique, strong internal magnetic fields can now be detected in red giant stars \citep{Fuller_2015}. This technique utilizes observations of mixed modes, oscillations that behave as pressure waves (p modes) near the stellar surface and gravity waves (g modes) in the stellar core \citep{1974A&A....36..107S,1975PASJ...27..237O,1977A&A....58...41A,2001MNRAS.328..601D,2004SoPh..220..137C,Dupret_2009}. Mixed modes have been used successfully to determine the evolutionary status of red giant stars \citep{Bedding_2011,stello_2013,mosser_2014}, and to measure their internal rotation rate \citep{Beck_2012,2012A&A...548A..10M,2014A&A...564A..27D,Deheuvels_2015}. Turbulent convection in the red giant envelope excites these modes, with part of the wave energy leaking through the evanescent region into the g mode cavity. The presence of a strong magnetic field in the core is able to alter the propagation of the gravity waves, trapping  mode energy in the core and effectively decreasing its visibility \citep[magnetic greenhouse effect,][]{Fuller_2015}. Theoretical predictions from \citet{Fuller_2015} have been verified by \citet{Stello_2016}, who applied the theory to a large sample of ascending red giant branch (RGB) stars observed by {\it Kepler} and found that the amplitudes of depressed dipole oscillation modes are consistent with nearly total wave energy loss in the core.

Red giants with strong internal magnetic fields (B$\gtrsim 10^5$ G) can thus be identified by the presence of suppressed oscillation modes in their oscillation spectra. These stars allow for a calculation of the minimum magnetic field $B_{c,{\rm min}}$ that must exist in the core. The amplitude of the suppressed oscillation modes depends on the amount of coupling between the p mode and g mode cavities, which is regulated by the extent of the evanescent region and depends on the angular degree, $\ell$, of the mode. Dipolar ($\ell=1$) modes have maximum coupling and therefore show the largest amplitude depression in the presence of strong core magnetic fields. \cite{Stello_2016} showed that strong magnetic fields are present in roughly 50\% of RGB stars with $M \gtrsim 1.5 \, M_\odot$, but are very rare in stars with $M \lesssim 1.1 \, M_\odot$. They interpreted this dichotomy as an effect of MS core dynamo-generated magnetic fields, which are generated in the convective cores of $M \gtrsim 1.1 \, M_\odot$ stars.

We start by extending the analyses of \citet{Fuller_2015} and \citet{Stello_2016}, providing additional analysis of the generation, evolution, and detectability of internal stellar magnetic fields. We find that quadrupole ($\ell=2$) modes should exhibit detectable suppression in red giants with magnetic cores, which could be used to validate the theory of \cite{Fuller_2015}. Moreover, we find that core helium-burning clump stars with magnetic cores will exhibit highly suppressed dipole modes, and should be easily detectable in {\it Kepler} data. Moderate field strengths ($B \gtrsim 10^4$-$10^5 \, {\rm G}$) are sufficient for dipole mode suppression in clump stars, provided those fields exist within the stably stratified shell around the convective He-burning core.

Next, we examine the generation and evolution of magnetic fields in stellar cores. Although strong ($B \! > \! 10^5 \, {\rm G}$) magnetic fields are likely a common outcome of core dynamos, the fields are confined within the mass coordinates of the convective region, which can help explain the rising incidence of magnetic fields in stars with $1.1 \, M_\odot \! \lesssim \! M \! \lesssim \! 1.5 \, M_\odot$ due to the larger convective core of the higher mass stars. We find it difficult to predict whether the core magnetic fields will survive through core He-burning and into the white dwarf stage of evolution, as magnetic fields could be altered, amplified, or erased by convective He-burning. However, we outline how observations of suppressed dipole modes in clump stars can distinguish these possibilities and provide a clear picture of internal magnetic field evolution throughout stellar evolution. Finally, we examine connections with magnetic fields in compact remnants and other types of stellar pulsators. We show that core dynamo-generated fields could be responsible for the strong magnetic fields at the surfaces of some white dwarfs and neutron stars. Core fields may also affect or suppress pulsations in some slowly pulsating B type stars, $\gamma$-Doradus stars, and subdwarf-B stars. 

In Section~\ref{visibility}, we calculate theoretical predictions for the visibility of suppressed dipolar and quadrupolar oscillation modes during the red giant   and the red clump   phases. Section~\ref{dynamo} examines the theory of magnetic field generation and evolution, and how this relates to observations of RGB stars, while Section~\ref{clump} extends this analysis to clump stars. In Section~\ref{rgb_implications} we focus on some implications of our results for RGB stars. We discuss connections with magnetic fields in compact remnants in Section~\ref{remnants} and other types of pulsators in Section~\ref{pulsators}, before concluding in Section~\ref{conclusion}.


\section{Dipole and Quadrupole Mode Visibility}\label{visibility}

\subsection{Red Giant Branch}
\label{rgb}


We now utilize the same method presented in \citet{Fuller_2015} to calculate expected visibilities of both dipole and quadrupole modes in stars with $1.25 \, M_\odot \leq M \leq 3 \, M_\odot$. The ratio of suppressed mode power to normal mode power is
\begin{equation}
\label{eqn:vsup}
\frac{V_{\rm sup}^2}{V_{\rm norm}^2} = \bigg[ 1 + \Delta \nu \,\tau \,T^2 \bigg]^{-1} \, .
\end{equation}
Here, $\Delta \nu$ is the large frequency separation, $\tau$ is the radial mode lifetime, and $T$ is the wave transmission coefficient through the evanescent zone. The value of $T$ can be calculated via
\begin{equation}
\label{eqn:T}
T  = \exp \bigg[ - \int^{r_2}_{r_1} dr \sqrt{ - \frac{ \big( L_\ell^2 - \omega^2 \big) \big(N^2 - \omega^2 \big) }{v_s^2 \omega^2} } \bigg] \, .
\end{equation}
Here, $r_1$ and $r_2$ are the lower and upper boundaries of the evanescent zone, $L_\ell^2 = l(l+1)v_s^2/r^2$ is the Lamb frequency squared, $N$ is the Brunt-Vaisala frequency, $\omega$ is the angular wave frequency, and $v_s$ is the sound speed. We calculate $\Delta \nu$ and the frequency of maximum power $\nu_{\rm max}$ using the scaling relations of  \citet{Kjeldsen:1995} with solar reference values from \citet{Huber_2011}.

We construct stellar models using MESA \citep[Modules for Experiments in Stellar Evolution, release 7456,][]{Paxton_2010,Paxton_2013,Paxton_2015} evolving them from the zero age main sequence to the end of core He-burning. The models are non-rotating and adopt the OPAL opacity tables \citep{Iglesias:96} and an initial metallicity of $Z=0.02$ with a mixture taken from \citet{Asplund:2005}.  
Convective regions have been calculated using the mixing-length theory (MLT) with $\alpha_{\rm MLT}=2.0$. The boundaries of convective regions are determined according to the Schwarzschild criterion. Exponentially decaying overshoot at the convective boundaries is included with a mixing parameter $f=0.018$ \citep{2000A&A...360..952H,Paxton_2010}. We include red giant mass-loss using the prescription of \citet{Reimers:1975} with $\eta=0.5$. The inlist used to calculate the models is provided in the Appendix.
  
Figure~\ref{fig:visibility} shows our predictions for reduced mode power, $(V_{\rm sup}/V_{\rm norm})^2$, as a function of $\nu_{\rm max}$ for stars of various masses as they evolve up the RGB. To first order, the reduced mode power is very similar for stars of different mass. The largest differences occur at low frequencies ($\nu_{\rm max} \lesssim 50 \, \mu {\rm Hz}$), where we predict more massive stars to show lower dipole mode visibilities. However, we caution that clump stars (see Section \ref{clump}) may be difficult to distinguish from RGB stars at these low frequencies.

Figure~\ref{fig:visibility} also shows predictions for the reduced power of suppressed quadrupole oscillation modes. Quadrupole modes are expected to exhibit significantly less suppression at all values of $\nu_{\rm max}$. However, for stars low on the RGB ($\nu_{\rm max} \gtrsim 150 \, \mu {\rm Hz}$), we find $(V_{\rm sup}/V_{\rm norm})^2 \lesssim 0.5 $ for quadrupole modes, i.e., substantial mode suppression is expected for quadrupole modes in addition to dipole modes. We predict this quadrupole mode suppression can easily be measured in existing {\it Kepler} data, and can be used to test the magnetic greenhouse hypothesis.

  \begin{figure*}[!t]
  \centering
  \begin{minipage}[t]{0.506\textwidth}
    \includegraphics[width=\textwidth]{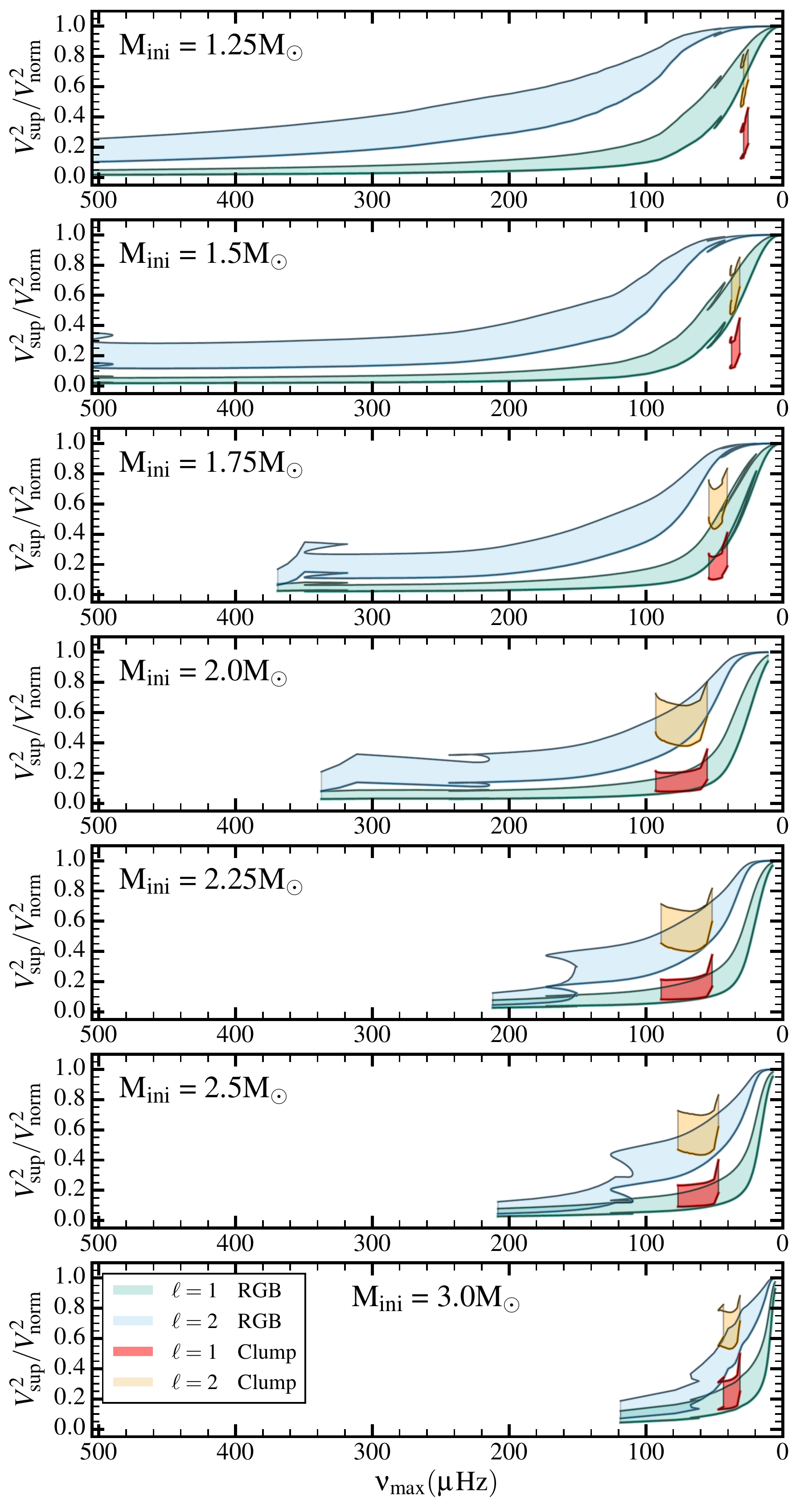}
  \end{minipage}
  \hfill
  \begin{minipage}[t]{0.483\textwidth}
    \includegraphics[width=\textwidth]{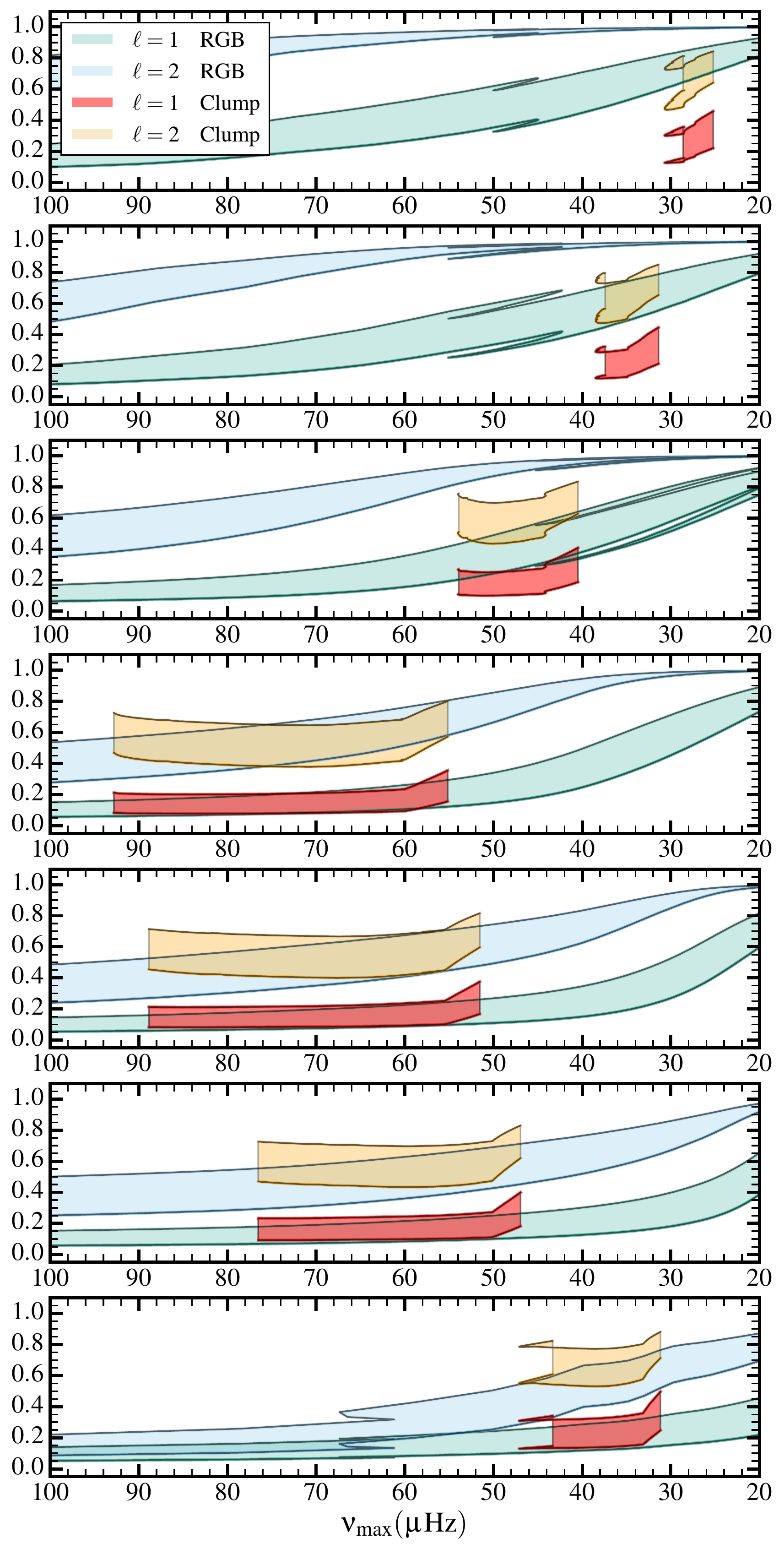}
  \end{minipage} \caption{\label{fig:visibility} Bands of visibility for suppressed $\ell=1$ and $\ell=2$ modes as a function of $\nu_{\rm max}$ for stars of different initial mass during H-shell burning (green/blue bands) and during He-core burning (Clump, red/orange bands). The upper and lower boundaries for the visibility bands correspond to values of radial mode lifetime $\tau=$ 10 and 30 days respectively, consistent with \citet{Dupret_2009,Corsaro_2015}. The right panel is a zoom of the left panel near the clump. Red giant branch calculations are shown after the H central mass fraction is $ X < 0.001$ and before the star reaches the tip of the RGB. Clump models are shown when the central mass fraction of He is $0.05 < Y < 0.9$.%
}
\end{figure*}


\subsection{Red Clump}
\label{clump}

In red giants, mixed modes are generally observable when the evanescent region separating the core and envelope is quite narrow, which occurs for RGB stars below the luminosity bump \citep{Dupret_2009,Grosjean_2014}. As stars evolve further up the RGB and expand, the evanescent region thickens, and mixed modes become undetectable. The mixed modes become visible again after the star has contracted and settled onto the clump for the core He burning phase. 

The degree of mode suppression due to the magnetic greenhouse effect in red giants behaves in a very similar fashion to mixed mode visibility, because both effects require wave energy to tunnel into the core. Therefore, dipole modes will have strongly suppressed visibilities in clump stars provided that sufficiently strong magnetic fields exist in their radiative cores. Figure~\ref{fig:visibility} shows the predicted power of dipole and quadrupole modes for clump stars of various masses. We predict typical suppressed powers of $0.1 \lesssim (V_{\rm sup}/V_{\rm norm})^2 \lesssim 0.4$ for suppressed dipole modes in clump stars, with more massive stars exhibiting slightly lower mode visibility. Quadrupole modes are expected to have $0.4 \lesssim (V_{\rm sup}/V_{\rm norm})^2 \lesssim 0.8$. 

For stars with $M \lesssim 2.0 \, M_\odot$, dipole modes will have smaller visibilities in clump stars than they will for RGB stars of the same mass and $\nu_{\rm max}$. Therefore, low-mass clump stars with suppressed modes may stick out by virtue of exhibiting exceptionally low dipole mode visibilities for stars with $\nu_{\rm max} \approx 25-50 \, \mu {\rm Hz}$. These stars may be visible as the group of stars with $V^2 \approx 0.4$ and $\nu_{\rm max} \approx 40 \, \mu {\rm Hz}$ in Figure~2 of \cite{Fuller_2015}. We caution that this region of $\nu_{\rm max}$ and visibility space is also inhabited by more massive stars ($M \gtrsim 2 \, M_\odot$) with suppressed modes that are ascending the RGB, although these two populations can be distinguished by their different masses. In larger samples, the existence or absence of a large population of stars with $V^2 \approx 0.2-0.4$ and $\nu_{\rm max} \approx 25-50 \, \mu {\rm Hz}$ will indicate whether magnetic mode suppression commonly operates within clump stars of $M \lesssim 2.0 \, M_\odot$. Secondary clump stars (with $M \gtrsim 2 \, M_\odot$) with suppressed modes will have similar mode visibilities to their RGB counterparts at the same $\nu_{\rm max}$, and distinguishing RGB from clump stars may be difficult for more massive stars.

\section{Features of main sequence dynamo-generated fields}
\label{dynamo}

\subsection{Operation of Main-Sequence Dynamo}\label{msdynamo}

The MS of stars more massive than about 1.1$\mso$ is characterized by the presence of a convective core,
which is expected to host a magnetic dynamo. Dynamo action converts a fraction of the kinetic energy 
of the convective motions into magnetic energy, with magnetic fields sustained against dissipation  \citep[see e.g.,][]{Brandenburg_2005}. The amplitude and scale of the generated magnetic field depends on the relative importance of rotation, which is usually quantified by the Rossby number, $Ro$. The Rossby number is the ratio between inertial and Coriolis forces, which is quantified by the ratio between the local rotation period and the convective eddy turnover timescale, $Ro = P_{\rm rot}/(2 t_{\rm con})$. 
Efficient dynamo action is expected for $Ro \lesssim 1$. In this case, the field is expected to be large scale and with an amplitude corresponding to equipartition between the magnetic energy density ($B^2/8\pi$) and the kinetic energy density in the flow ($\rho v^2_{\rm con}/2$), where $v_{\rm con}$ is an RMS convective velocity.

Typical values for the surface rotation periods of A stars are short \citep[about 1 day, see e.g.][]{Zorec_2012}. 
Moreover, asteroseismic observations of slowly rotating A stars suggest these stars are nearly rigidly rotating \citep{Kurtz_2014}, and very little angular momentum is lost during the MS for stars above $1.3\mso$ \citep[Kraft break, see e.g.][]{1967ApJ...150..551K,2013ApJ...776...67V}. 
Convective turnover timescales within the core are generally larger, with $t_{\rm con} = 2 \alpha H_P/v_{\rm con} \sim 1 \, {\rm month}$ for a $1.5 \, M_\odot$ model. Here, $\alpha H_P$ is the mixing length and $H_P$ is a pressure scale height, and we evaluate $v_{\rm con}$ from mixing length theory. In the convective core, $v_{\rm con} \approx (F/\rho)^{1/3}$, where $F$ is the energy flux carried by convection.
We therefore expect $Ro \ll 1$ in the cores of stars above $1.1$-$1.3\mso$. 
We can then estimate magnetic field strengths in the convective cores of MS stars assuming that the dynamo creates a magnetic field of equipartition strength,
\begin{equation}
\label{eqn:Beq}
B_{\rm MS} = \sqrt{ 4 \pi \rho v_{\rm con}^2} \, .
\end{equation}
We find typical field strengths of $B_{\rm MS} \sim 10^4$-$10^5 \, {\rm G}$. Note however that smaller scale, smaller amplitude magnetic fields can still be generated in the absence of rapid rotation, so sizable magnetic fluxes might well be ubiquitous in stellar convective cores. 

These estimates are supported by magneto-hydrodynamics simulations of the central regions of MS stars. For example, for a $2\mso$ A-type star rotating with periods of 28 and 7 days, \citet{Brun_2005}
show dynamo action with magnetic fields reaching a considerable fraction of equipartition ($B\approx10^4$-$10^5$G). 
Note also that the generation of equipartition and even super-equipartition magnetic fields with peak strengths above 1 megagauss (MG)  has been shown in MHD simulations of core convection in massive B-type stars \cite{2013PhDT.......388A}.

\begin{figure}[!tb]
\begin{center}
\includegraphics[width=1\columnwidth]{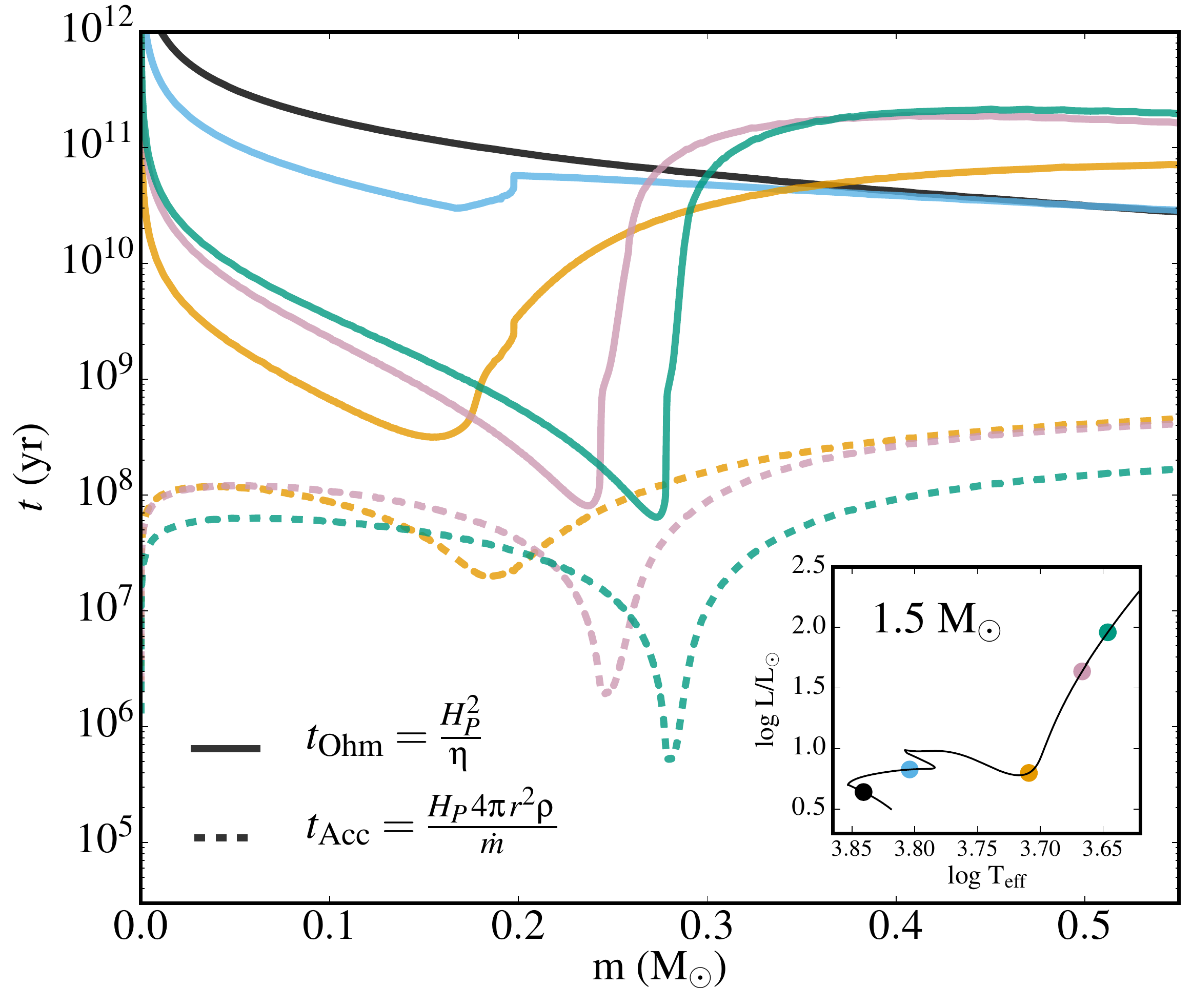}
\caption{\label{fig:timescales}
Ohmic diffusion timescale, $t_{\rm Ohm}$, and mass accretion timescale onto the helium core, $t_{\rm Acc}$, for a $1.5 \, M_\odot$ stellar model evolving from the MS up the RGB. Values are shown for the inner $0.55 \, M_\odot$ of the model. The evolutionary stages for which the profiles are extracted are identified by the colored dots on the evolutionary track (see the inset). The accretion timescale can only be calculated after the end of the main sequence, when a helium core is present. At all phases of evolution and at all radii, $t_{\rm Ohm} \gg t_{\rm Acc}$, and so a core dynamo-generated magnetic field is unable to significantly diffuse outward.%
}
\end{center}
\end{figure}

\subsection{Diffusion vs Advection}
\label{time}

We assume that at the end of the MS a magnetic field is present below the maximum extent of the convective core during core H-burning.
The first important timescale is the Ohmic timescale $t_{\rm Ohm}= H_{\rm P}^2/\eta$, the time it takes a stable magnetic field in a radiative region to diffuse across a pressure scale height. This timescale is usually quite long, due to the small values of the magnetic diffusivity $\eta$ in stellar plasmas. Figure~\ref{fig:timescales} shows that in the core of a $1.5\mso$ star evolving from the MS to the RGB, $t_{\rm Ohm}$ varies between $10^8$-$10^{12}$ years. Therefore  magnetic fields present in the stellar core at the end of the MS are frozen in their Lagrangian mass coordinate. Note that  $t_{\rm Ohm}$ does not depend on the amplitude or geometry of the magnetic field, but only on the local value of the magnetic diffusivity. The magnetic diffusivity is the inverse of the electrical conductivity,  which in RGB stars has to be calculated carefully as certain regions are partially/fully degenerate. Moving from non-degenerate, to partially and fully degenerate regions, we calculate the magnetic diffusivity according to \cite{1968dms..book.....S}, \cite{1987ApJ...313..284W} and \cite{1984MNRAS.209..511N} respectively, applying a smooth interpolation in the transition regions.

\citet{MacGregor_2003} discuss the possibility that magnetic buoyancy instabilities during the MS can bring small magnetized fibrils to the stellar surface. However, the inclusion of realistic compositional gradients seems to disfavor this scenario, increasing considerably the timescales of magnetic buoyancy \citep{MacDonald_2004}.

The other important timescale is the H-shell burning timescale. As a star moves from the end of its MS to the RGB phase, the ashes of H-shell burning increase the size of its He core. We can write the timescale of this process as $t_{\rm Acc} = H_{\rm P} 4\pi r^2 \rho / \dot{m}$, where $r$ is the local radial coordinate and $\dot{m}$ is the He accretion rate. If $t_{\rm Acc} \! < \! t_{\rm Ohm}$, then the magnetic field can be buried below the He raining from the H-shell. Figure~\ref{fig:timescales} shows that in a $1.5\mso$ star this is always the case. As the He core grows substantially during the sub-giant/early RGB phases, magnetic fields from MS core convection can be efficiently buried below the H-shell, the location where the waves are most sensitive to the magnetic greenhouse effect (see e.g. the $1.25\mso$ model in Fig.~\ref{fig:DipoleHist}).






  

\subsection{Extent of Magnetic Fields and Magnetic Mode Suppression}
\label{rgb}

The strong magnetic fields created by a core dynamo will be mostly confined to the region occupied by the convective core during the MS. Simulations show that the field strength drops off rapidly in overlying regions (\citealt{Featherstone_2009}, Augustson et al. 2015), and we showed in Section \ref{time} that the field will not be able to substantially diffuse outward during the lifetime of the star. 

\begin{figure}[!tb]
\begin{center}
\includegraphics[width=1\columnwidth]{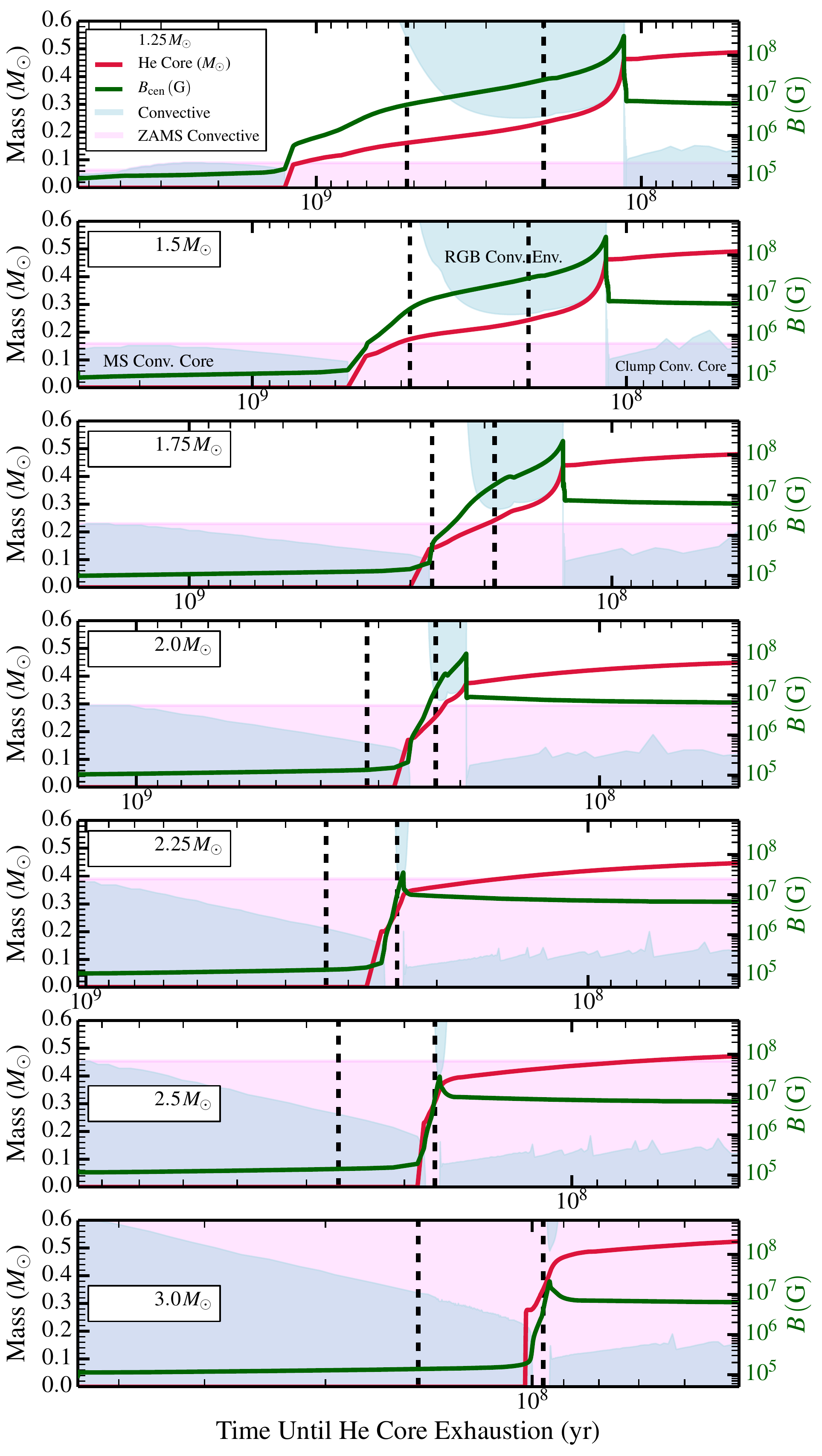}
\caption{\label{fig:DipoleHist}
Kippenhahn diagrams of the inner regions of intermediate-mass stars, shown from the main sequence to the red clump phases of evolution. Blue shaded regions are convective zones, labeled in the second panel. Pink shaded regions indicate mass coordinates that were within the main sequence convective zone and contain strong core-dynamo-generated fields. The red line is the location of the H-burning shell during post-main-sequence evolution. The green line estimates the central magnetic field strength, $B_{\rm cen}$, calculated via equation \ref{eqn:Bcen}. Dashed lines denote the location of $\nu_{\rm max} = 500 \, \mu{\rm Hz}$ (left line) and $\nu_{\rm max} = 50 \, \mu{\rm Hz}$ (right line). Mixed modes and suppressed modes are generally observable when the star lies between the vertical dashed lines. Magnetic suppression is most likely when the red line lies within a pink shaded region, i.e., it is most likely to be observed for $M \gtrsim 1.5 \, M_\odot$, in accordance with the results of \cite{Stello_2016}. For the same evolutionary stage, the frequency of maximum power shifts to lower values for stars of higher mass, such that $\nu_{\rm max} = 500 \, \mu{\rm Hz}$ corresponds to the main sequence for our most massive models.%
}
\end{center}
\end{figure}

The size of the convective core changes significantly during MS evolution, as shown in Figure~\ref{fig:DipoleHist}. Here, we have used the Schwarzschild criterion to determine the extent of the convective region, as appropriate for stars in this mass range \citep{Moore_2015}. 
Generally, the mass contained within the convective core decreases as stars approach core hydrogen depletion. However, this does not imply that the extent of strongly magnetized regions decreases, because any region which is convective at some point during the MS may contain strong fields previously deposited by the dynamo action. This assertion is consistent with simulations \citep{Featherstone_2009} that indicate that convective core dynamos do not destroy overlying magnetic fields, and with the detection of magnetic fields long after the termination of MS dynamos \citep{Stello_2016}.

During post-MS evolution, we therefore expect that strong fields will exist only within regions that were convective at some point during the MS, indicated by the pink shaded regions in Figure~\ref{fig:DipoleHist}. These fields are most likely to lead to magnetic suppression on the RGB if they exist at the mass coordinate of the hydrogen burning shell (red line in Figure~\ref{fig:DipoleHist}). Moreover, the suppression will only be evident for stars in the sub-giant/lower RGB phase of evolution, approximately in the range $50 \, \mu {\rm Hz} \lesssim \nu_{\rm max} \lesssim 500 \, \mu {\rm Hz}$, which is when the coupling between p- and g- modes is strongest (shown by vertical dashed lines in Figure~\ref{fig:DipoleHist}). Hence, suppression is most likely to be observed when the red line lies within a pink shaded region and between the vertical dashed lines. Magnetic suppression is unlikely to be observed in stars with $M \lesssim 1.5 \, M_\odot$ because the H-burning shell lies above the magnetized regions during the RGB. Magnetic suppression is much more common in stars with $M \gtrsim 1.5 \, M_\odot$, for which the H-burning shell lies within magnetized regions on the lower RGB. We propose that this feature of stellar evolution helps account for the sharp rise in magnetic suppression for masses $M \gtrsim 1.5 \, M_\odot$ found by \cite{Stello_2016}.

\subsection{Magnetic Field Strength and Structure}
\label{fieldstruc}

To calculate plausible magnetic field strengths in the cores of red giant stars, we assume that the MS convective core dynamo creates a magnetic field of equipartition strength.
Using Eq.~\ref{eqn:Beq} we find typical field strengths of $B_{\rm MS} \sim 2 \! \times \! 10^5 \, {\rm G}$; again due to the very long diffusion timescales, we expect these fields to be confined to the maximal extent of the core convective region. As in \citet{Fuller_2015} we then assume that magnetic flux is conserved as the star evolves into a red giant, such that
\begin{equation}
\label{eqn:Brgb}
B_{\rm RG} = B_{\rm MS} \bigg( \frac{r_{\rm MS}}{r_{\rm RG}} \bigg)^2 \, .
\end{equation}
Here, the red giant magnetic field $B_{\rm RG}$ at a mass coordinate $m$ is calculated from the corresponding MS field $B_{\rm MS}$ at the moment when the convective core has its largest extent. The radius $r_{\rm MS}$ is the radial coordinate of the mass shell at this point on the MS, while the radius $r_{\rm RG}$ is the radial coordinate of the mass shell when the star is on the clump. 

The post-MS core magnetic field can also be approximated from the global properties of the star without detailed stellar models, as shown in Appendix \ref{Bcenap}. We find that a reasonable estimate of a red giant's central magnetic field strength, $B_{\rm cen}$, is
\begin{equation}
\label{eqn:Bcen}
B_{\rm cen} \sim L_{\rm MS}^{1/3} M_{\rm c,MS}^{-2/9} \rho_{\rm c,MS}^{-5/18} \rho_{\rm c,RG}^{2/3} \, .
\end{equation}
Here, $L_{\rm MS}$, $M_{\rm c,MS}$, and $\rho_{\rm c,MS}$, are the MS luminosity, convective core mass, and central density, respectively, while $\rho_{\rm c,RG}$ is the post-MS central density. Figure~\ref{fig:DipoleHist} plots $B_{\rm cen}$ in our stellar models. Central magnetic fields of $B_{\rm cen} \sim 10^5 \, {\rm G}$ are expected during the MS, while field strengths of $B_{\rm cen} \sim 10^7 \, {\rm G}$ are expected on the clump if the fields survive to this phase of evolution. In general, field strengths of $B_{\rm RG} > 10^6 \, {\rm G}$ can be expected near the centers of red giant stars.

The spatial structure of a MS dynamo-generated field has important implications for the suppression of oscillation modes during the red giant phase.
The angular structure of the field will affect magnetic mode splitting and wave scattering, which could potentially be used to constrain the magnetic field structure within red giants.

While the dynamo is active, we expect the magnetic field to vary on horizontal scales comparable to those of convective eddies, as seen in simulations \citep{Featherstone_2009}. The largest convective eddies at the outer edge of the convective core have length scales of $\sim \! H_P \sim \! r_c$, where $r_c$ is the radius of the convective core. Since these magnetic structures can be long-lived, we expect them to be mostly frozen in to the core when it becomes radiative at the end of the MS, as long as the global field structure remains stable. Therefore, we expect dominant fluctuations in the angular structure of the field to occur on length scales of $\sim \! r_c$, although smaller structures may also exist due to smaller eddies within the convective flow.

To extrapolate these structures into the red giant phase, we assume they remain frozen in place as the core contracts. During this process, the structures shrink by a factor of $r_{\rm RG}/r_{\rm MS}$.
Hence, we expect the field to vary on horizontal length scales up to the radial coordinate $r$ in red giant cores. The Ohmic diffusion time across structures of this size is large (see Section~\ref{time}), such that they are not able to be significantly smoothed out within the lifetime of the star.

Horizontal variations break the spherical symmetry of the field and affect its interaction with oscillations. In principle, one could envision a field structure that is tangled only on very small scales $l$, such that its surface resembles the dimpled surface of a cantaloupe and appears nearly spherically symmetric to an incoming wave. For this to happen, the length scale $l$ must be smaller than the {\it radial} wavelength of incoming waves such that $k_r l < 1$. However, we argued above that the field will vary on length scales $r$, and \citet{Fuller_2015} showed that $k_r r \gg 1$ within the cores of red giants for observable oscillation modes. Therefore, realistic field configurations can always break the spherical symmetry of the background and scatter incoming waves into high angular wave numbers $k_\perp$ such that the magnetic greenhouse effect operates as described in \citet{Fuller_2015}.

\section{Magnetic Mode Suppression in Red Clump Stars}
\label{clump}

The understanding of magnetic mode suppression presented in Section \ref{rgb} can be used to make predictions for magnetic mode suppression in red clump stars. We expect that mode suppression will occur if strong fields created by convective core dynamos can exist within the radiative regions of clump stars.

\begin{figure*}[!tbp]
  \centering
  \begin{minipage}[t]{0.48\textwidth}
    \includegraphics[width=\textwidth]{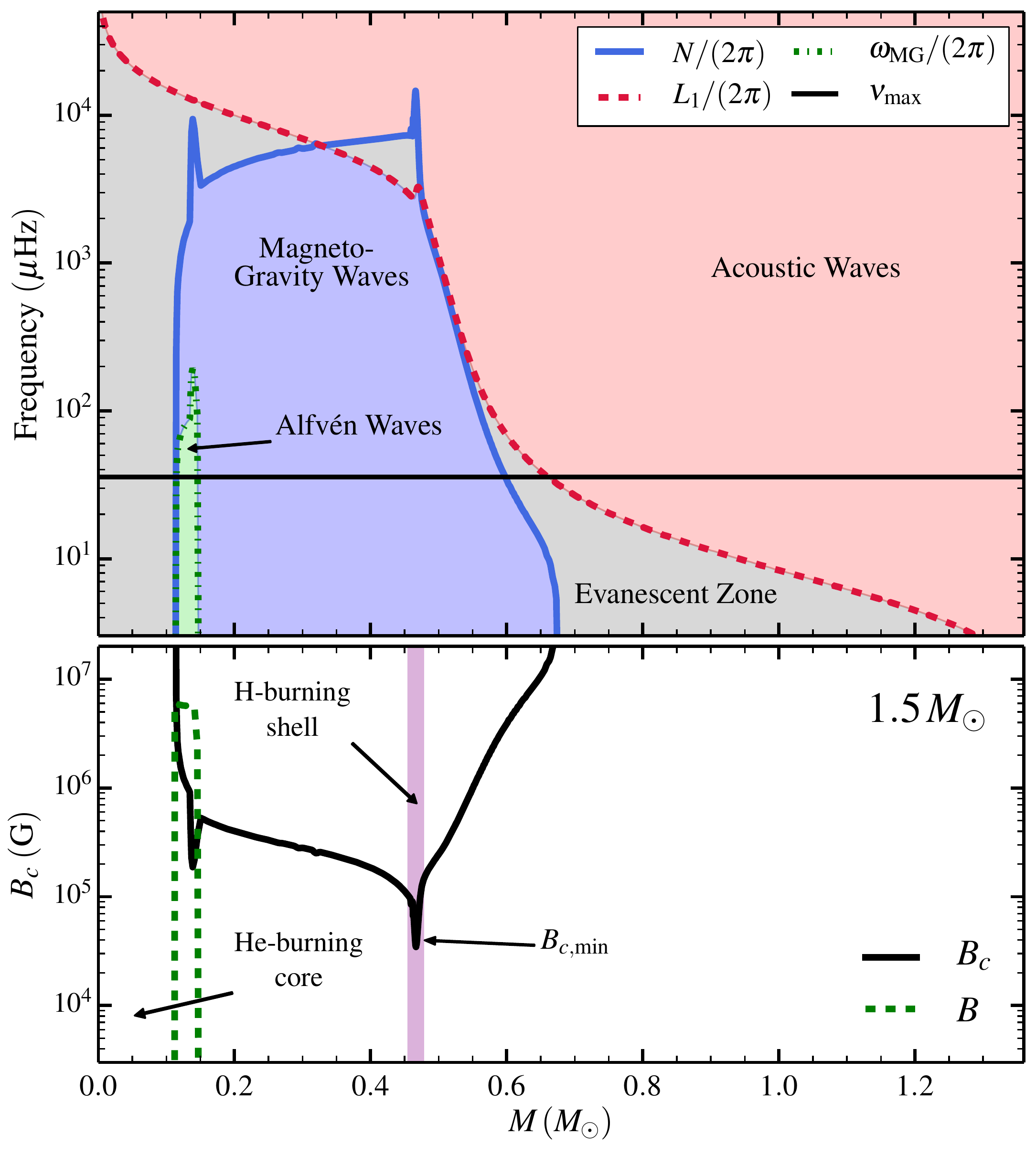}
  \end{minipage}
  \hfill
  \begin{minipage}[t]{0.49\textwidth}
    \includegraphics[width=\textwidth]{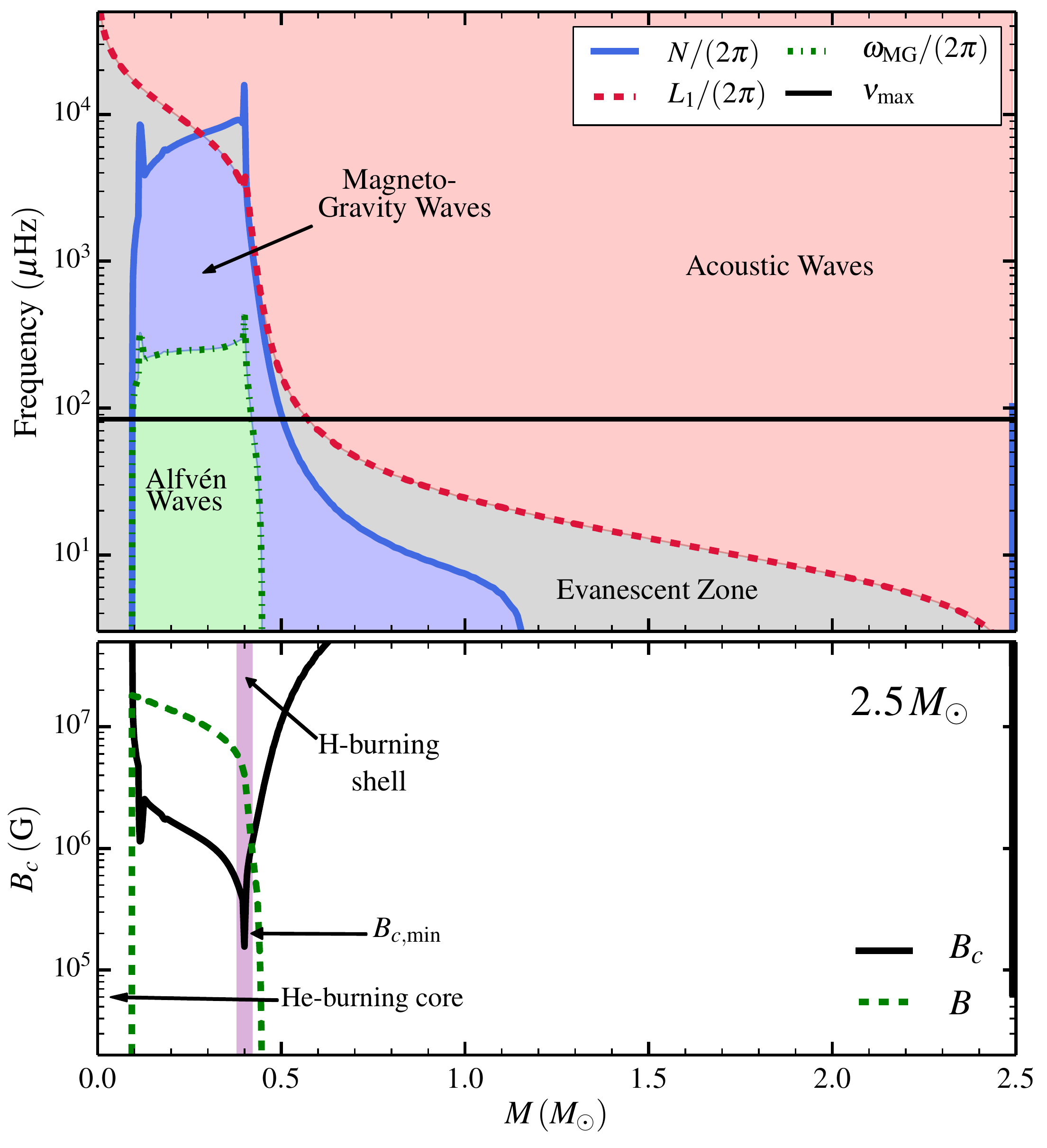}
  \end{minipage}
  \caption{\label{fig:ClumpProp} Left: Propagation diagram for a $1.5 \, M_\odot$ model on the clump. 
  Colored regions are labeled by the types of waves they support. Waves near the frequency of maximum power ($\nu_{\rm max}$, black horizontal line) are acoustic waves in the envelope, magneto-gravity waves in the outer core, and Alfv\'en waves in strongly magnetized regions.
  The magnetized areas (green regions) correspond to stably stratified regions which supported a convective dynamo on the main sequence. 
   The bottom panel shows the critical magnetic field strength $B_c$ required for magnetic suppression (solid black line) and predicted magnetic field in the radiative region outside the convective core calculated  assuming magnetic flux conservation from equipartition fields generated via a main sequence dynamo (green dashed line). Magnetic suppression is expected to occur if $B \! > \! B_c$, which in this model occurs just outside of the convective He-burning core.  Right: Same as left panel, but for a $2.5 \, M_\odot$ model. For this model, the magnetized regions extend beyond the H-burning shell on the clump (pink vertical line).}
\end{figure*}

Figure~\ref{fig:ClumpProp}  shows propagation diagrams of clump stars with zero age main sequence masses of $1.5 \, M_\odot$ and $2.5 \, M_\odot$. 
Again we assumed that the field has been created by an equipartition convective core dynamo during the MS  (eq.~\ref{eqn:Beq}). We then estimate core magnetic fields in the radiative regions of clump stars assuming magnetic flux conservation (eq.~\ref{eqn:Brgb}). As shown in Figure~\ref{fig:DipoleHist} and in the bottom panels of Figure~\ref{fig:ClumpProp}, magnetic fields of $B_{\rm RG} > 10^6 \, {\rm G}$ may exist in the cores of clump stars.

\begin{figure}[!tb]
\begin{center}
\includegraphics[width=1\columnwidth]{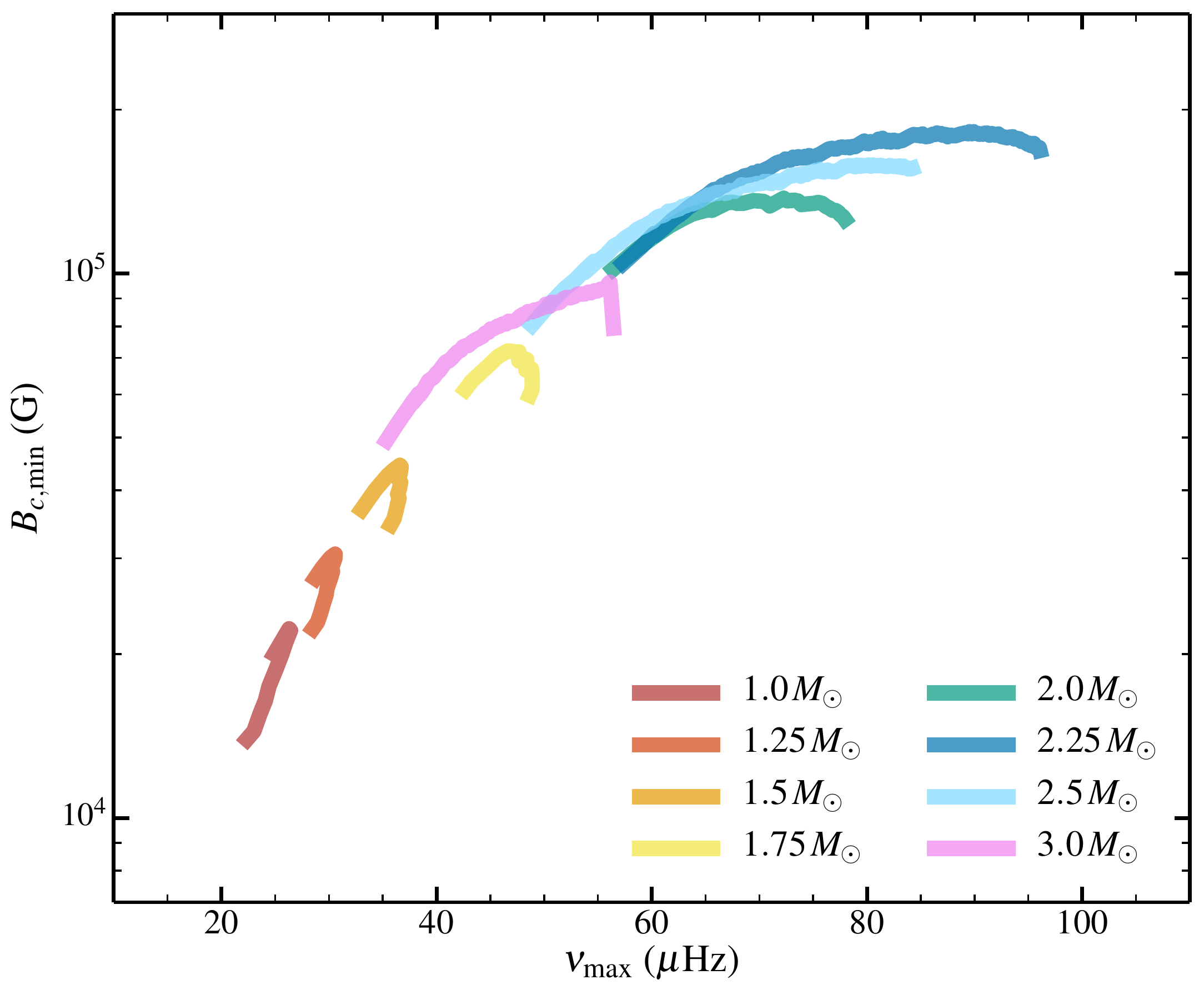}
\caption{\label{fig:BcClump}
Minimum core magnetic field strength $B_{c,{\rm min}}$ required for magnetic suppression on the clump. The tracks are plotted during core He-burning when the central mass fraction of He is $0.05 < Y < 0.9$. Low-mass stars ($M<2.0 \, M_\odot$) begin clump evolution at the bottom of the tracks and evolve counter-clockwise along the tracks. Higher mass stars ($M>2.0 \, M_\odot$) begin at the right-hand side of the tracks and evolve toward the left as the stellar envelope expands and the value of $\nu_{\rm max}$ decreases.%
}
\end{center}
\end{figure}

Figure~\ref{fig:ClumpProp} also shows the magneto-gravity frequency $\omega_{\rm MG}$ \citep{Fuller_2015}
\begin{equation}
\label{eqn:omegaMG}
\omega_{\rm MG} = \bigg[ \frac{ 2 B_{\rm RG}^2 N^2}{\pi \rho r^2} \bigg]^{1/4} \, .
\end{equation}
Magnetic suppression is expected if $\nu_{\rm max} < \omega_{\rm MG}/(2 \pi)$ at some point in the radiative region. This is equivalent to the requirement that $B_{\rm RG} > B_c$, where $B_c$ is the critical magnetic field strength
\begin{equation}
\label{eqn:Bc}
B_c = \sqrt{ \frac{ \pi \rho}{2} } \frac{\omega^2 r}{N} \, ,
\end{equation}
evaluated at angular wave frequencies $\omega = 2 \pi \nu_{\rm max}$. Figure~\ref{fig:BcClump} shows the minimum field strength required for mode suppression, $B_{c,{\rm min}}$, evaluated at the peak in $N$ at the H-burning shell, for stars on the red clump. In general, fields of $10^4-2 \! \times \! 10^5 \, {\rm G}$ are sufficient for mode suppression. 

Figure~\ref{fig:ClumpProp}  indicates that magnetic fields strong enough to cause mode suppression may commonly exist within the radiative cores of clump stars. In stars of $M \lesssim 2.25 \, M_\odot$, these fields will only exist below the H-burning shell, however, they are likely still strong enough to cause magnetic mode suppression. In stars of $M \gtrsim 2.25 \, M_\odot$, the strong fields extend beyond the H-burning shell and will very likely lead to mode suppression. These conclusions are not sensitive to the spike in $N$ just above the convective core (which occurs because of the composition gradient between convective core and radiative region) and are not sensitive to mixing processes at the convective core boundary. 

Inspection of Figures~\ref{fig:DipoleHist} and \ref{fig:ClumpProp} demonstrates another interesting feature: in stars of $M \lesssim 1.4 M_\odot$, the mass of the convective core on the clump is larger than its maximal extent on the MS. Therefore, we do not expect strong fields in the radiative regions of low-mass $M \lesssim 1.4 M_\odot$ clump stars, and we predict that low-mass clump stars will rarely exhibit oscillation mode suppression. Hence, mode suppression on the clump may be similar to that measured for RGB stars measured by \cite{Stello_2016}: mode suppression may be rare for stars with $M \lesssim 1.3 M_\odot$ (occurring in less than $\sim \! 10 \%$ of stars), and common for stars $M \gtrsim 1.5 M_\odot$ (occurring in greater than $\sim \! 50 \%$ of stars).


%
%

\subsection{Convective Magnetic Field Destruction}

The predictions outlined above assumed that magnetic fields generated by core dynamos during the MS are preserved through red giant evolution. While this is likely a good assumption for magnetic fields in radiative regions (see Section \ref{time}), it may not hold in regions of the star that become convective during post-MS evolution. In fact, convection may destroy dynamo-generated fields during later phases of evolution, especially if the newly formed convective regions are slowly rotating (as measured in many red giant cores, \citealt{Mosser_2012}), such that a large scale dynamo does not operate.

We emphasize that it is not clear whether convection will destroy pre-existing stable field configurations. If the convective energy density, $\epsilon_{\rm con}$, is larger than the magnetic energy density, $\epsilon_{\rm mag}$, the convective motions may not be constrained by the magnetic fields. In this case, the convection could scramble the pre-existing field into an unstable configuration. In the case where $\epsilon_{\rm con} < \epsilon_{\rm mag}$, convective motions could be confined along the field lines, such that the field remains nearly unaltered. Nonetheless, the convective diffusivity may still be able to erode the stable fields on relatively short timescales.

\begin{figure}[!tb]
\begin{center}
\includegraphics[width=1\columnwidth]{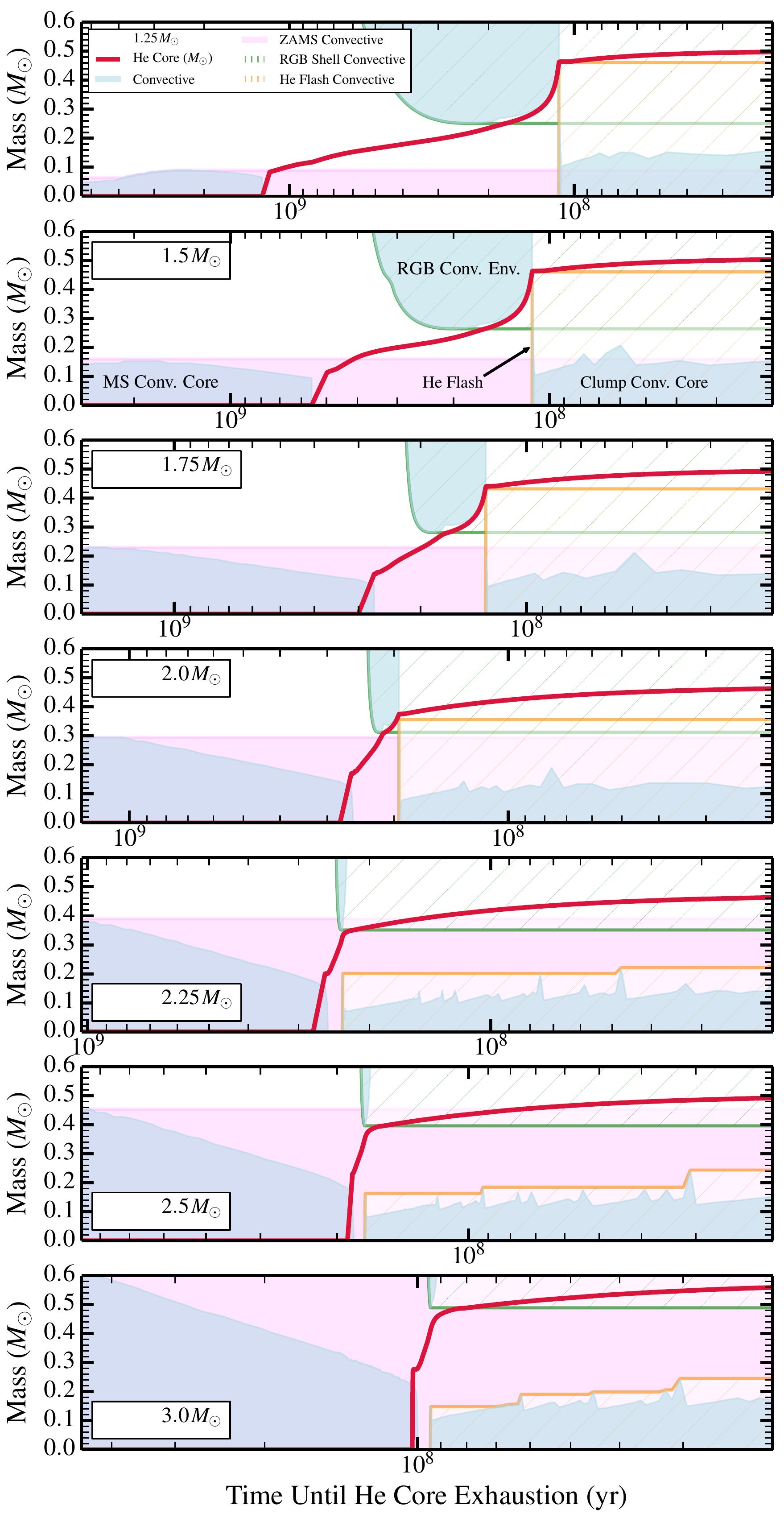}
\caption{\label{fig:DipoleHistConv}
Same as Figure~\ref{fig:DipoleHist}, but we have also marked mass coordinates that were in the convective envelope during the RGB evolution (green hatched areas) and regions that were convective during He shell flashes and/or He core burning (orange hatched regions). If convection during RGB/clump evolution destroys strong magnetic fields created during the main sequence, we should not expect magnetic suppression to occur in stars that undergo He shell flashes, i.e., stars of $M \lesssim 2.1 \, M_\odot$. However, magnetic suppression may still be possible in secondary clump stars, i.e., stars of $M \gtrsim 2.1 \, M_\odot$.%
}
\end{center}
\end{figure}

There are three convective phases during red giant evolution that may destroy MS dynamo-generated fields. First, the convective envelope can extend below mass coordinates of $\approx 0.3 \, M_\odot$ (see Figure~\ref{fig:DipoleHistConv}) during the first dredge-up when the star ascends the RGB. In these regions, we find that $\epsilon_{\rm con} \! > \! \epsilon_{\rm mag}$, evaluating $\epsilon_{\rm mag}$ using equations \ref{eqn:Beq} and \ref{eqn:Brgb}. It is therefore possible that any fields at mass coordinates above the deepest extent of the convective envelope are destroyed.

Second, very vigorous convection develops during He flashes in stars of $M \lesssim 2 \, M_\odot$ \citep{Bildsten_2011}. We find that $\epsilon_{\rm con} \gg \epsilon_{\rm mag}$ during He flashes, such that the short-lived convection can likely destroy pre-existing fields. Figure~\ref{fig:DipoleHistConv} shows that these flashes induce convection in all mass coordinates below $ \sim \! 0.4 \, M_\odot$ in low-mass stars. We find that stars with $M \lesssim 2.1 \, M_\odot$ have evolved such that {\it all} mass coordinates  of the radiative core on the clump were convective at some point of prior red giant evolution. 

Third, the He-burning convective core in clump stars could consume fields within this region. This process is not necessarily relevant for oscillation mode suppression, which relies on strong magnetic fields within radiative regions. However, if destruction of the field within the convective core destabilizes the field in overlying radiative regions, core convection could still have an impact. We find that $\epsilon_{\rm con} \sim \epsilon_{\rm mag}$ within the convective cores of our stellar models on the clump. Thus, it is not clear whether MS dynamo-generated fields will be destroyed during core He burning. 

The discussion above allows for a prediction about the occurrence of magnetic mode suppression in clump stars. If convection in red giants destroys previously existing fields, we expect oscillation mode suppression to only occur in relatively massive ($M \gtrsim 2.1 \, M_\odot$) secondary clump stars. Only these massive stars contain radiative regions that were convective on the MS (and may contain strong dynamo-generated fields) but which were not convective during prior evolution on the RGB or during He flashes. On the other hand, if convection in red giants is able to generate stable magnetic fields via a dynamo process, we expect oscillation mode suppression to be common in clump stars of all masses.

The observation (or lack thereof) of oscillation mode suppression in clump stars will therefore provide great understanding of magnetic field evolution in stellar interiors. If mode suppression is common in clump stars with masses as small as $\sim \! 1.5 \, M_\odot$, this would indicate that magnetic fields are robust, and are able to survive through post-MS convective phases. If mode suppression is common in clump stars with masses as small as $\sim \! 1.0 \, M_\odot$, it would indicate post-MS convection can frequently generate strong and long-lived magnetic fields. If mode suppression only occurs in clump stars with $M \gtrsim 2.1 \, M_\odot$, this would indicate that post-MS convection generally destroys pre-existing fields. 

We caution that these results can be somewhat influenced by the size of the convective core,  which may not be accurately calculated by stellar evolution codes such as MESA because of the unknown extent of mixing induced by, e.g., convective overshoot.
Indeed, asteroseismic studies of clump stars \citep{montalban_2013,stello_2013,mosser_2014,bossini_2015,constantino_2015} and sub-dwarf B stars \citep{vangrootel_2010a,vangrootel_2010b,charpinet_2011,Schindler_2015} indicate that the convective core is somewhat larger than predicted by stellar evolution codes, even using optimistic overshooting prescriptions. MS convective cores may also be somewhat larger and show evidence for enhanced mixing \citep{moravveji_2015}.  Our calculations of mode visibilities and field strengths $B_c$ are not strongly affected, but inferences based on the extent of the convective core become less certain.

\section{Implications for RGB Stars}\label{rgb_implications}
\subsection{Incidence Rate of Strong Magnetic Fields in Red Giant Stars}
\citet{Stello_2016} determined the incidence of strong internal magnetic fields as a function of stellar mass in ascending RGB stars. Stars that did not have convective cores during the MS ($M<1.1\mso$) do not show suppressed dipole modes, while red giants with $1.1 \, M_\odot \lesssim M \lesssim 1.5 \, M_\odot$ exhibit an increasing incidence in mode suppression rate with mass. In Section \ref{rgb} we discussed how the extension of the MS convective core could help create this trend. 

Another important consideration is the increasing efficiency of dynamo action with rotation rate. Stars with relatively rapidly rotating convective cores (small Rossby numbers $Ro$, see Sec.~\ref{msdynamo}) would generate strong core fields, while stars with slowly rotating cores might not produce strong and/or stable internal magnetic fields. Indeed, several recent studies of dynamo action in young solar type stars \citep{Vidotto_2014,See_2015,Folsom_2016} have shown that their dynamos produce field strengths that saturate for $Ro \lesssim 0.1$, and which scale as $B \propto Ro^{-1}-Ro^{-1.5}$ for $Ro \gtrsim 0.1$. Since the core convective turnover time scales for stars with $1.1 \, M_\odot \lesssim M \lesssim 1.5 \, M_\odot$ are roughly one month, we expect substantially reduced field strengths in the cores of stars which are magnetically braked to rotation periods of $P \gtrsim 3 \, {\rm d}$ by the end of the MS. Indeed, stars of $M\lesssim 1.3 \, M_\odot$ lie below the Kraft break and can likely be spun down to such rates by the end of the MS. \citep[see][]{VanSaders_2013,McQuillan_2014}. Their dynamos may then shut down, and this may help explain the increasing prevalence of strong magnetic fields with stellar mass in the $1.1$-$1.5 \, M_\odot$ mass range.


However, this does not explain why the incidence rate of depressed dipole modes appears to saturate at $\approx 50\%$ in red giants with $M \gtrsim 1.5 \mso$. One possibility is that, at the end of the MS and its associated dynamo, the magnetic field is able to evolve into a stable configuration only in $\approx 50\%$ of the cases. The evolution of magnetic fields into stable magnetic configurations has been studied \citep{Braithwaite_2006}, but it is difficult to make detailed predictions as the outcome depends on the complex initial conditions of the magnetic configuration left by turbulent convection, in particular magnetic energy and  helicity \citep{Braithwaite_2008}. Moreover, these theoretical calculations do not include a compositional stratification, which instead might play an important role in confining the magnetic field after dynamo action ceases in the convective cores of stars. 

We also note that magnetic field occurrence rates from \cite{Stello_2016} could be influenced by uncertainties of order $\Delta M/M \! \sim \! 5-10 \%$ in the asteroseismic masses inferred from scaling relations. For instance, a sharp jump in magnetic field occurrence rate as a function of mass will be smoothed out by roughly $\Delta M \! \sim \! 0.1 \, M_\odot$. Additionally, due to a steep initial-mass function and short lifetime on the lower RGB \citep[see][]{Lloyd_2011,Lloyd_2013}, the high-mass end of the distribution ($M \gtrsim 1.6 \, M_\odot$) can be somewhat contaminated by outliers from the larger sample of low-mass stars. Thus, the suppression rate of $50-60\%$ found by \cite{Stello_2016} at the high-mass end could in fact be a lower limit, as some low-mass stars without strong core fields may dilute the high-mass sample. This effect should be quantified by more detailed population/statistical studies.

\subsection{Magnetic Splitting of Mixed Modes}

Some red giants may contain magnetic fields in their core that are not strong enough to produce oscillation mode suppression via the magnetic greenhouse effect. Instead, these stars may exhibit magnetically split oscillation modes. In the limit that $B \ll B_c$, the mode frequency perturbation is \citep{Unno_1989}
\begin{equation}
\label{eqn:dommag}
\frac{\delta \omega_{\rm M}}{\omega} = \frac{1}{8 \pi \omega^2 I} \int d V | \delta {\bf B} |^2 \, .
\end{equation}
Here, $I$ is the mode inertia,  $\delta {\bf B}$ is the perturbation to the magnetic field, and the integral is taken over the volume of the star. The degree of mode splitting is proportional to $B^2$, and is thus very small for $B \ll B_c$.

In Appendix A, we show that the magnetic mode splitting for modes of frequency $\nu$ is approximately
\begin{equation}
\label{eqn:dfmag}
\delta \nu_{\rm M} \sim \frac{\sqrt{\ell(\ell+1)}}{8 \pi} \Delta \nu_{\rm g} \int^{R_{\rm g}}_0 \frac{N}{\omega} \bigg(\frac{B_r}{B_c}\bigg)^2 \frac{dr}{r} \, .
\end{equation}
Here, $\Delta \nu_{\rm g}$ is the g mode frequency spacing between mixed modes of angular degree $\ell$, $B_r$ is the radial component of the field, $B_c$ is the critical field strength (equation \ref{eqn:Bc}), and the integral is taken over the g mode cavity

To estimate the level of magnetic splitting expected, we evaluate equation \ref{eqn:dfmag} for the red giant model shown in Figure~S1 of \cite{Fuller_2015}, but with a field strength weaker by a factor of 10, such that $B_r \simeq B_c$ at the H-burning shell and $B_r \ll B_c$ everywhere else. This is roughly the maximum field strength that could exist without oscillation mode suppression, and this particular model has a central field strength of $B_r \approx 7 \times 10^5 \, {\rm G}$.  In this case, we find $\delta \nu_{\rm M} \approx 2 \, \mu {\rm Hz}$, whereas $\Delta \nu_{\rm g} \approx 1 \, \mu{\rm Hz}$. Magnetic splitting may therefore be comparable to the g mode period spacing in RGB stars on the verge of oscillation mode suppression. The magnetic splitting may also be comparable to rotational splitting. In these stars (of which ``Droopy", KIC 8561221, \citealt{Garcia_2014}, may be an example), magnetic splitting may complicate the interpretation of g mode period spacing and rotational splitting. We abstain from a more thorough analysis of mode splitting in these stars, as it depends on both field geometry and core rotation rate. However, in most ``normal" stars with slightly weaker fields such that $B_r \ll B_c$ everywhere, we expect $\delta \nu_{\rm M}$ to be much smaller than rotational splitting and is likely undetectable.

\subsection{Angular Momentum Transport}

Several asteroseismic studies \citep{Beck_2012,Mosser_2012,Deheuvels_2014,Deheuvels_2015,Dimauro:2016} have measured the core rotation rates of RGB/clump stars. The relatively slow core rotation rates indicate strong angular momentum transport mechanisms are at work \citep{Cantiello_2014}, coupling the radiative cores with the convective envelope. The strong magnetic fields frequently found in the cores of RGB stars \citep{Stello_2016} may play an important role in this process \citep{2014ApJ...793..123M,Kissin_2015}. However our work suggests that strong magnetic fields are restricted to mass coordinates of RGB stars that were convective on the MS. For stars of $M \lesssim 1.5 \, M_\odot$, the strong fields are restricted to the He core, and cannot directly couple the core with the envelope. Since the majority of the sample of \cite{Mosser_2012} has $M \lesssim 1.5 \, M_\odot$, core-dynamo-generated fields cannot solely account for slow core rotation.

More importantly, the sample of stars with measured core rotation rates are mutually exclusive from the sample of stars with strong core magnetic fields. The reason is that mixed dipole modes are used to measure the core rotation rates, but these modes are highly suppressed/absent in stars with magnetic cores. In order for large-scale magnetic fields to account for the measured core rotation rates, they must extend from the He core to the outer radiative core, and they must be weak enough that they do not suppress dipole oscillation modes. Unfortunately, neither the current study nor the measurements of \cite{Stello_2016} can provide useful constraints on the existence of such fields.

\section{Implications for Remnants}
\label{remnants}

\subsection{Magnetic Fields in White Dwarfs}

Many white dwarfs (WDs) exhibit strong ($B \gtrsim 3 \, {\rm MG}$) surface magnetic fields. The exact fraction of WDs which have strong fields is debated but is of order 10\% \citep{Hollands_2015}. Moreover, there is a dearth of magnetic WDs with field strengths below $10 \, {\rm MG}$, and the magnetic WDs are systematically more massive \citep{Ferrario_2015B}.

Our work suggests that some of the strong fields observed at WD surfaces could be the remnants of MS core-dynamo-generated fields. Equipartition fields of $\sim \! 2 \times 10^5 \, {\rm G}$ generated by a core dynamo would evolve into fields of $\sim 2 \times 10^7 \, {\rm G}$ if their flux is conserved from the MS to the WD phase. Thus, the field strengths inferred for core dynamo-generated fields are squarely within the observed distribution of WD surface fields \citep{Ferrario_2015B,Ferrario_2015A}.

Figure~\ref{fig:DipoleHist} shows that the core convective region is restricted to mass coordinates below $\approx 0.5 \, M_\odot$ for stars of $M \lesssim 2.5 M_\odot$. Using the MS-WD mass relation of \citet{Renedo_2010,2015ApJ...815...63A}, these low-mass stars account for the majority of WDs, whose mass distribution peaks near $0.6 \, M_\odot$ \citep{Rebassa-Mansergas_2015}. Since the ohmic diffusion time is very long in WDs \citep{Ferrario_2015B}, we expect that core dynamo-generated fields are unlikely to be visible at the surface of most WDs. 

However, in stars of $M \gtrsim 3 \, M_\odot$, the convective core extends to mass coordinates larger than $0.6 \, M_\odot$.  In these stars, the entire mass of the WD descendant lies within the mass coordinate occupied by the MS convective core. Therefore, strongly magnetized regions may extend all the way to the WD surface where they can be observed. Stars over $M \gtrsim 3 \, M_\odot$ produce WDs of $M_{\rm WD} \gtrsim 0.7 \, M_\odot$, similar to the typical masses of magnetic WDs \citep{Ferrario_2015B}. Thus, some magnetic WDs may be magnetized due to MS convective core dynamos, which could partially explain why magnetic WDs are more massive on average. However, it remains possible that magnetic WDs are the descendants of magnetic Ap stars, or that they are formed through WD mergers. 

The interesting corollary to this discussion is that strong magnetic fields may exist within the interiors of many WDs even though the fields are not visible at the surface. Indeed, based on the MS-WD mass relation, a significant fraction of WDs originate from progenitors with $M \gtrsim 1.5 \, M_\odot$ in which core dynamos operate and produce strong magnetic fields in over 50\% of RGB stellar cores \citep{Stello_2016}. We therefore speculate that many (perhaps the majority of) WDs could contain strong ($B \gtrsim 10^6 \, {\rm G}$) magnetic fields which are confined within the stellar interior and not detectable at the surface even as they cool. This is because the WD cooling timescale is shorter than its magnetic diffusion timescale $t_{\rm Ohm} \approx 10^{11}$ yrs \citep{Cumming_2002}. These magnetic fields may have very important implications for WD evolution, and for the outcome of WD mergers.

\subsubsection{Helium-Core White Dwarfs}

Helium-core white dwarfs (He WDs) typically have masses in the range $0.15 \, M_\odot \lesssim M \lesssim 0.4 \, M_\odot$ and are formed when a companion star strips the hydrogen envelope of the He WD progenitor as it ascends the RGB. He WDs are essentially the naked cores of the RGB stars analyzed in \cite{Fuller_2015} and \cite{Stello_2016}. Unless internal magnetic fields are somehow destroyed by envelope mass loss, we expect some He WDs to exhibit surface fields of $B \gtrsim 10^5 \, {\rm G}$. As far as we are aware, strong magnetic fields have not yet been observed at the surfaces of any He WDs, even though they may be detectable.

Predicting the fraction of He WDs that will exhibit strong surface fields is not straightforward, as it depends both on the progenitor mass and the He WD mass. For instance, the $1.75 \, M_\odot$ model shown in Figure~\ref{fig:DipoleHist} has an MS convective core that extends to a mass coordinate of $\approx 0.24 \, M_\odot$. Therefore, its He WD descendant may only exhibit strong surface fields if its mass is $M_{\rm WD} \lesssim 0.24 \, M_\odot$, otherwise the fields may remain buried. We encourage searches for magnetic fields in He WDs, as their detection would allow further characterization of the strong fields inferred to exist within red giant cores.

\subsection{Magnetic Fields in Neutron Stars}

Since the observations of \cite{Stello_2016} show that core-dynamo-generated magnetic fields frequently survive well into RGB evolution in low-mass stars, it is possible that these fields are also long-lived in massive stars that spawn neutron stars upon their death. We find typical equipartition field strengths of $B \sim 10^6 \, {\rm G}$ in the MS convective core of $M \sim 12 \, M_\odot$ neutron star progenitors. Flux conservation of the field within the inner $1.4 \, M_\odot$ (which has a radius of $\sim \! 0.5 \, R_\odot$ on the MS) to a neutron star radius of $12 \, {\rm km}$ would lead to neutron star surface field strengths of $\sim \! 10^{15} \, {\rm G}$, i.e., magnetar field strengths.  

The magnetar birth rate is highly uncertain, with plausible estimates in the range of \citep{keane_1998,mereghetti_2015} $5-50\%$ of the neutron star birth rate. It is therefore possible that stable magnetic fields in the cores of massive stars are just as common as in low-mass stars, leading to magnetar birth rates of order $50 \%$ of the neutron star birth rate.  
If magnetar birth rates turn out to be smaller, it may indicate that post-MS core convective phases (He, C, Ne, O, or Si burning) destroy MS core dynamo-generated fields and prevent magnetar formation, as we have hypothesized to happen in low-mass stars. 
An absence of oscillation mode suppression in intermediate-mass clump stars would support this hypothesis. It also remains possible that most magnetars are the descendants of magnetic OB stars, or that their fields are generated during a proto-neutron star dynamo \cite{1992ApJ...392L...9D}.

\section{Implications for Other Asteroseismic Targets}
\label{pulsators}


Subdwarf B (sdB) stars are essentially the naked He cores of red clump stars, with masses of $M \simeq 0.47 \, M_\odot$ \citep{fontaine_2012} and thin H envelopes of $M_{\rm H} \sim 10^{-3} \, M_\odot$. They provide an opportunity to constrain much of the physics discussed in this paper, since magnetic fields which would be confined to the cores of clump stars could be visible at the surfaces of sdB stars. If strong fields can be detected in sdB stars, it may indicate that magnetic oscillation mode suppression will occur in clump stars. However, \cite{Landstreet_2012} observe no evidence for strong magnetic fields at the surface of any known sdB star, and find that strong fields occur in less than a few percent of sdB stars. 

The lack of strong magnetic fields at the surfaces of sdB stars may have two causes. First, sdB stars evolve primarily from low-mass ($M \lesssim 2.25 \, M_\odot$) stars which have been stripped of their H envelope just prior to the He flash \citep{heber_2009}. Their MS progenitors had convective cores of $M_{\rm core} \lesssim 0.4 \, M_\odot$ (see Figure~\ref{fig:DipoleHist}), and therefore any dynamo-generated fields are likely confined to the interiors of sdB stars and are not detectable at their surfaces. Second, sdB stars evolved through a He flash phase, and strong large-scale fields may have been wiped out by convection during that time (see Figure~\ref{fig:DipoleHistConv}). Thus, the apparent absence of strong fields at sdB surfaces is not altogether surprising. However, some small fraction of sdBs likely evolve from low-mass magnetic Ap stars, and therefore we may expect to see strong fields (if they are not wiped out by He flashes) at the surfaces of a small percentage of sdB stars. Additional observations, coupled with sdB population synthesis calculations, will be needed to reach a robust conclusion.

Many sdB stars pulsate in p modes (periods of $\sim$minutes) and/or g modes (periods of $\sim$hours). Their pulsations may be used to study magnetic mode alteration in sdBs with strong internal fields. A propagation diagram for an sdB star is shown in Figure~\ref{fig:sdBProp}. We find that a magnetic field of $B\sim 10^5 \, {\rm G}$ near the He-H transition (located at $r/R \approx 0.35$ in Figure~\ref{fig:sdBProp}) is sufficient to strongly alter a typical sdB g mode with a frequency of $\nu = 100 \, \mu {\rm Hz}$. Fields as small as $B\sim 10^3 \, {\rm G}$ near the surface (at $r/R \approx 0.95$) could also create magnetic alteration. Even smaller fields can strongly alter lower frequency modes, although we caution that our conclusions are somewhat sensitive to the mass of the H-envelope and the operation of diffusive/mixing processes. Unlike mixed modes in red giants, g modes in sdB stars are not separated from the surface by a thick evanescent region, and therefore magnetically altered magneto-gravity modes could be detectable at the surface. Therefore, we strongly encourage detailed analyses of the g mode spectra of pulsating sdB stars, as the pulsations may carry information about strong sub-surface fields.

  
Since magnetic mode suppression is relatively common in red giant pulsators, it is possible that it operates (but has not been recognized) in other types of pulsators as well. In MS stars, g modes are most vulnerable to magnetic alteration because smaller field strengths are required to suppress lower frequency oscillations (see equation \ref{eqn:Bc}). The magnetic greenhouse effect (as described by \citealt{Fuller_2015}) may not operate in the same manner, but strong magnetic fields may still spread the power of oscillation modes into a broad range of spherical harmonics $\ell$ and therefore reduce their visibility. 

In $\gamma$-Doradus stars, it is possible that strong magnetic fields located just outside the convective core may inhibit the development of large amplitude pulsations in some stars (Fig.~\ref{fig:GdorProp}). In particular, stars passing through the $\gamma$-Dor instability strip at the end of their MS evolution may contain strong magnetic fields that have been deposited in the radiative region around the shrinking convective core. We find that the approximate critical field strength required to inhibit pulsations with a frequency of $\nu = 10 \, \mu {\rm Hz}$ in a $1.6 \, M_\odot$ star passing through the $\gamma$-Dor instability strip is $B_c \approx 10^5 \, {\rm G}$, although the precise value depends somewhat on the value of $N$ (and therefore the mixing processes at work) just outside the core. This field strength is lower than the equipartition fields that could have been deposited during previous MS evolution (we find $B_{\rm eq} \approx 2 \times 10^5 \, {\rm G}$), and it is therefore possible that strong magnetic fields inhibit or alter $\gamma$-Dor pulsation modes in some stars within the instability strip. We also find that more modest fields of $B \sim 10^{3} \, {\rm G}$ are capable of altering g modes near the surface of the star where $\rho$ is much smaller. Therefore, magnetic Ap/Fp stars in the $\gamma$-Dor instability strip may exhibit magnetically altered/suppressed g modes. 

Slowly pulsating B (SPB) stars also exhibit g mode pulsations which could be altered by strong internal magnetic fields (Fig.~\ref{fig:SPBProp}). We find very similar field characteristics to those in $\gamma$-Dor stars could alter the g modes in a $5 \, M\odot$ SPB model. A field of $\sim 10^5 \, {\rm G}$ just outside the convective core, or a field of $\sim 10^{3} \, {\rm G}$ near the surface would suffice to alter modes of $\nu_{\rm g} = 10 \, \mu {\rm Hz}$. The SPB star $\zeta$ Cas \citep{Neiner_2003,Briquet_2016} exhibits a surface field slightly weaker than this, and presents an interesting opportunity to study the magnetic field-pulsation interaction. \cite{Hasan_2015} find a field strength of $\approx 10^5 \, {\rm G}$ near the core will produce a $\sim 1 \%$ frequency splitting in SPB g modes. This estimate may be appropriate if $B \! < \! B_c$ everywhere, but more sophisticated (non-perturbative) calculations are needed for stars with $B \! > \! B_c$ somewhere in their interiors.

\section{Conclusions and Future Work}
\label{conclusion}

The main goal of this paper is to examine the implications of asteroseismic detections of strong internal magnetic fields in low-mass red giant stars via suppressed dipole oscillation modes \citep{Fuller_2015,Stello_2016}, and to generate predictions, extrapolations, and guidelines for future work. 

First, we examined the visibility of suppressed oscillation modes in stars with strong core magnetic fields. Quadrupole modes are predicted to exhibit small (but detectable) suppression which can be used to check the validity of the existing theory and observational techniques. Dipole mode suppression should be easily detectable in clump stars in addition to stars ascending the RGB. For stars with $M \lesssim 2 \, M_\odot$, clump stars are expected to show lower dipole mode visibility than ascending RGB stars at the same $\nu_{\rm max}$, and therefore magnetic clump and RGB stars might be distinguished from one another.

Next, we investigated the evolution of magnetic fields created by MS convective core dynamos during post-MS evolution. Magnetic diffusion timescales within the red giant core are generally longer than stellar evolution time scales. Therefore, the fields will be frozen into the mass coordinates at which they form (i.e., mass coordinates within the MS convective core), and will not migrate outward during red giant evolution. In stars with $M \lesssim 1.5 \, M_\odot$, the H-burning shell lies above the extent of the MS convective core during evolution on the lower RGB. Therefore, strong fields are not expected to be present at the location of the H-burning shell in these stars, which may account for the low incidence of dipole suppression in ascending RGB stars with $M \leq 1.5 \, M_\odot$ \cite{Stello_2016}. The opposite is true for stars with $M \gtrsim 1.5 \, M_\odot$, accounting for their much larger observed dipole mode suppression rate. Additionally, more rapid rotation in the higher mass stars may generate stronger core fields and contribute to their higher suppression rates.

We then examined the possibility of magnetic mode suppression for clump stars. Field strengths in the range of $2$-$20 \! \times \! 10^4 \, {\rm G}$ at the H-burning shell are required for mode suppression in clump stars, or fields a few times stronger in the radiative He above the convective He-burning core. However, it is not clear whether such fields will persist into the clump phase, as it is possible that they are erased/altered by vigorous convection during He flashes. We anticipate three possibilities for dipole mode suppression in clump stars:\\
1. The incidence of dipole mode suppression is similar to that for RGB stars, indicating core magnetic fields are not strongly affected by convective He burning phases. \\
2. Dipole mode suppression is common even in stars with $M \lesssim 1.3 \, M_\odot$, indicating that He flashes tend to generate strong core magnetic fields. \\
3. Dipole mode suppression is common only in stars of $M \gtrsim 2.1 \, M_\odot$, indicating that He flashes tend to eliminate strong core magnetic fields. We tentatively advocate that this is the most likely scenario.


Finally, we discussed implications for the magnetic fields in white dwarfs (WDs), neutron stars (NSs), and pulsating stars. We suggest that strong fields may exist within the interiors of $50 \%$ or more of WDs, but are preferentially visible at the surfaces of massive WDs. Only the massive WD progenitors ($M \gtrsim 3 \, M_\odot$) had MS convective cores which encompassed the entire mass of the WD remnant, and hence only these stars could have generated strong MS fields capable of being observed at the surface of the WD. In NS progenitors, the MS convective core always extends to mass coordinates larger than the NS mass, and equipartition fields entail NS field strengths of $\sim \! 10^{15} \, {\rm G}$ if magnetic flux is conserved after the MS. Therefore, MS dynamos may be capable of creating the fields observed in magnetars if they survive into the NS phase and are not destroyed by subsequent convective core/shell burning phases. 

\section*{Acknowledgments} 

This paper was written collaboratively, on the web, using \href{https://www.authorea.com}{Authorea}. We thank Bill Paxton, Dennis Stello, Rafa Garcia, Jamie Lloyd and all the members of the SPIDER collaboration for helpful discussions and suggestions. JF acknowledges partial support from NSF under grant no. AST-1205732 and through a Lee DuBridge Fellowship at Caltech. This research was supported by the National Science Foundation under grant No. NSF PHY11- 25915 and AST 11-09174, and by NASA under TCAN grant No. NNX14AB53G.

\appendix

\section{Approximating the Core Magnetic Field Strength}
\label{Bcenap}

A rough estimate for the core magnetic field strength of a star with a convective MS core can be calculated as follows. The MS luminosity is efficiently carried by core convective motions, and implies an equipartition field strength of 
\begin{equation}
\label{eqn:Beq2}
B_{\rm eq} \simeq (4 \pi \rho)^{1/6} L^{1/3} r^{-2/3} \, ,
\end{equation}
where we have used $v_{\rm con} \simeq \big[L/(4 \pi \rho r^2)\big]^{1/3}$ to rewrite equation \ref{eqn:Beq}, and $L$ is the stellar luminosity. In general, this expression is a function of radius. At the edge of the convective core, we have
\begin{equation}
\label{eqn:rc}
r_{\rm c} = \bigg[ \frac{3 M_{\rm c}}{4 \pi {\bar \rho}_{\rm c}} \bigg]^{1/3} \, ,
\end{equation}
where $M_{\rm c}$ is the mass of the convective core and ${\bar \rho}_{\rm c}$ is its average density. Then we estimate within the convective core an approximate field strength of
\begin{equation}
\label{eqn:Beq3}
B_{\rm MS} \simeq 3^{-2/9} (4 \pi)^{7/18} \rho^{1/6} {\bar \rho}_{\rm c}^{\, 2/9} L^{1/3} M_{\rm c}^{-2/9} \, .
\end{equation}
Within the convective core, the density does not change greatly. Since equation \ref{eqn:Beq3} scales weakly with density, we can use the approximation $\rho \sim {\bar \rho}_{\rm c} \sim \rho_{\rm c}$ , where $\rho_{\rm c}$ is the central density. We can also drop the numerical prefactor which is of order unity. This expression should only be considered an order of magnitude approximation of the field strength within the convective core, since quantities such as $L$, $\rho$, and the enclosed mass are functions of radius.

Next, consider a sphere of density $\rho_{\rm c}$ near the center of the star. To conserve its mass, its radius evolves as $r_{\rm RG}/r_{\rm MS} = (\rho_{\rm c,MS}/\rho_{\rm c,RG})^{1/3}$. Then assuming magnetic flux conservation after the MS as given by equation \ref{eqn:Brgb}, we arrive at 
\begin{equation}
\label{eqn:Bcen2}
B_{\rm RG} \sim L_{\rm MS}^{1/3} M_{\rm c,MS}^{-2/9} \rho_{\rm c,MS}^{-5/18} \rho_{\rm c,RG}^{2/3} \, .
\end{equation}
This expression can be easily evaluated from basic stellar models without a detailed knowledge of their structure. The very weak scaling with $M_{\rm c,MS}$ and $\rho_{\rm c,MS}$ ensures that the approximations above are appropriate, and that only rough estimates of these quantities are needed to calculate a reasonable core field strength. Equation \ref{eqn:Bcen2} is a reasonable approximation for mass coordinates below the hydrogen burning shell whose density is not greatly different from the central density. It breaks down for mass coordinates above the hydrogen burning shell which have much lower densities.

\section{Magnetic Mode Splitting}
\label{magmode}

Here we provide an approximate formula to estimate the magnitude of magnetic splitting from equation \ref{eqn:dommag}. The perturbed magnetic field is
\begin{equation}
\label{eqn:dB}
\delta {\bf B} = {\boldsymbol \nabla} \times \big( {\boldsymbol \xi} \times {\bf B} \big) \, ,
\end{equation}
where ${\boldsymbol \xi}$ is the displacement produced by the oscillation mode. To simplify the calculation, we adopt an unphysical scenario of a magnetic field purely in the radial direction. In this scenario the splitting is independent of mode angular number $m$; the more realistic case of a dipolar field is treated in \cite{Unno_1989} (but beware of typos in equation 19.65) and \cite{Hasan_2015}, and can produce different splittings for different $m$ modes depending on the angle between the magnetic and rotation axes.

For a radially symmetric field, equation \ref{eqn:dB} reduces in the WKB limit to 
\begin{equation}
\delta {\boldsymbol B} \simeq i B_r k_r \boldsymbol{\xi}_\perp \, ,
\end{equation}
where $\boldsymbol{\xi}_\perp = \xi_\perp r \boldsymbol{\nabla} Y_{lm}$ is the horizontal wave displacement. Equation \ref{eqn:dommag} then yields
\begin{equation}
\label{eqn:dommag2}
\frac{\delta \omega_{\rm M}}{\omega} \simeq \frac{\ell(\ell+1)}{8 \pi \omega^2 } \frac{\int r^2 k_r^2 \xi_\perp^2 B_r^2 d r }{\int  \rho r^2 \Big[ \xi_r^2 + \ell(\ell+1) \xi_\perp^2 \Big] d r }\, .
\end{equation}
For a mode with most of its inertia in the g mode cavity, this can be written 
\begin{equation}
\label{eqn:dommag3}
\frac{\delta \omega_{\rm M}}{\omega} \sim \frac{1}{8} \frac{\int \rho r^2 \xi_\perp^2 (B_r/B_c)^2 d r }{\int \rho r^2 \xi_\perp^2 d r }\, ,
\end{equation}
where we have used the definition of $B_c$ from equation \ref{eqn:Bc}. Pressure dominated modes will exhibit slightly smaller magnetic splitting due to their inertia in the envelope.

In the WKB limit, the quantity $\rho r^2 \omega^2 v_{{\rm g},r} \xi_\perp^2$ (which represents an energy flux) is constant. Here $v_{{\rm g},r}$ is the gravity wave group velocity in the radial direction, $v_{{\rm g},r} = \omega^2 r/\sqrt{\ell(\ell+1)N^2}$. Then we have
\begin{equation}
\label{eqn:dommag4}
\frac{\delta \omega_{\rm M}}{\omega} \sim \frac{1}{8} \frac{\int v_{{\rm g},r}^{-1} (B_r/B_c)^2 d r }{\int v_{{\rm g},r}^{-1} d r }\, .
\end{equation}
The denominator is directly related to the g mode period spacing $\Delta P_{\rm g}$ (see equation 12 of \citealt{Chaplin_2013}),
\begin{equation}
\label{eqn:dpg}
\int v_{{\rm g},r}^{-1} dr = \frac{2 \pi^2}{\omega^2 \Delta P_{\rm g}} = \frac{1}{2 \Delta \nu_{\rm g}} \, ,
\end{equation}
and $\Delta \nu _{\rm g}$ is the associated frequency splitting. After inserting this expression into equation \ref{eqn:dommag4}, a little rearranging yields equation \ref{eqn:dfmag}. We emphasize that for a realistic field configuration, the frequency perturbation will depend on both $\ell$ and $m$ (allowing modes to be magnetically split), and may be different by a factor of a few.

\newpage

\section{MESA Inlist}
\label{inlist}

Here the inlist used to calculate the stellar evolution models discussed in the paper
\begin{verbatim}

&star_job
            
      change_lnPgas_flag = .true.
      new_lnPgas_flag = .true.
      pgstar_flag = .true.

/ ! end of star_job namelist

&controls
      
      !----------------------------------------  MAIN
       
      initial_mass = 1.5
      initial_z = 0.02
      use_Type2_opacities = .true.
      Zbase = 0.02  
      
      !----------------------------------------  WIND


      RGB_wind_scheme = 'Reimers'
      AGB_wind_scheme = 'Blocker'
      RGB_to_AGB_wind_switch = 1d-4
      Reimers_wind_eta = 5d-1  
      Blocker_wind_eta = 5d-1  

      !----------------------------------------  OVERSHOOTING

      overshoot_f_below_nonburn = 0.018
      overshoot_f_above_burn_h = 0.018
      overshoot_f_above_burn_he = 0.018


      !----------------------------------------  MISC

      photostep = 100
      profile_interval = 100
      max_num_profile_models = 100
      history_interval = 1
      terminal_cnt = 10
      write_header_frequency = 10
      max_number_backups = 50
      max_number_retries = 100
      max_timestep = 3.15d14  ! in seconds  

      !----------------------------------------  MESH      

      mesh_delta_coeff = 0.8
      varcontrol_target = 5.d-4
      
      !----------------------------------------  STOP WHEN

      xa_central_lower_limit_species(1) = 'he4'
      xa_central_lower_limit(1) = 0.05
      
/ ! end of controls namelist

&pgstar       

/ ! end of pgstar namelist

\end{verbatim}

\section{Propagation Diagrams}

It is informative to understand the approximate magnetic field strengths $B_c$ needed for dipole mode alteration in various types of stars observed to pulsate in g modes. Below, we present propagation diagrams for three types of g mode pulsators: sdB stars, $\gamma$-Dor stars, and slowly pulsating B stars.  The propagation diagrams illustrate some of the general similarity between various types of g mode pulsators. Although very different in terms of mass, evolutionary state, etc., each of these models contains a convective core surrounded by a radiative envelope that comprises the g mode cavity. An approximate rule of thumb is that field strengths of order $B \sim 10^5 \, {\rm G}$ just outside the convective core, or field strengths of order $B \sim 10^3 \, {\rm G}$ near the surface, are required for strong g mode alteration. The value of $B_c$ at the edge of the convective core should be interpreted cautiously because it depends on mixing/overshoot processes that may substantially alter the value of $N$ and therefore the value of $B_c$ at this location. Values of $B_c$ for different mode frequencies $\nu$ can be calculated using $B_c \propto \nu^2$. 

\begin{figure}[h!]
\begin{center}
\includegraphics[width=0.45\columnwidth]{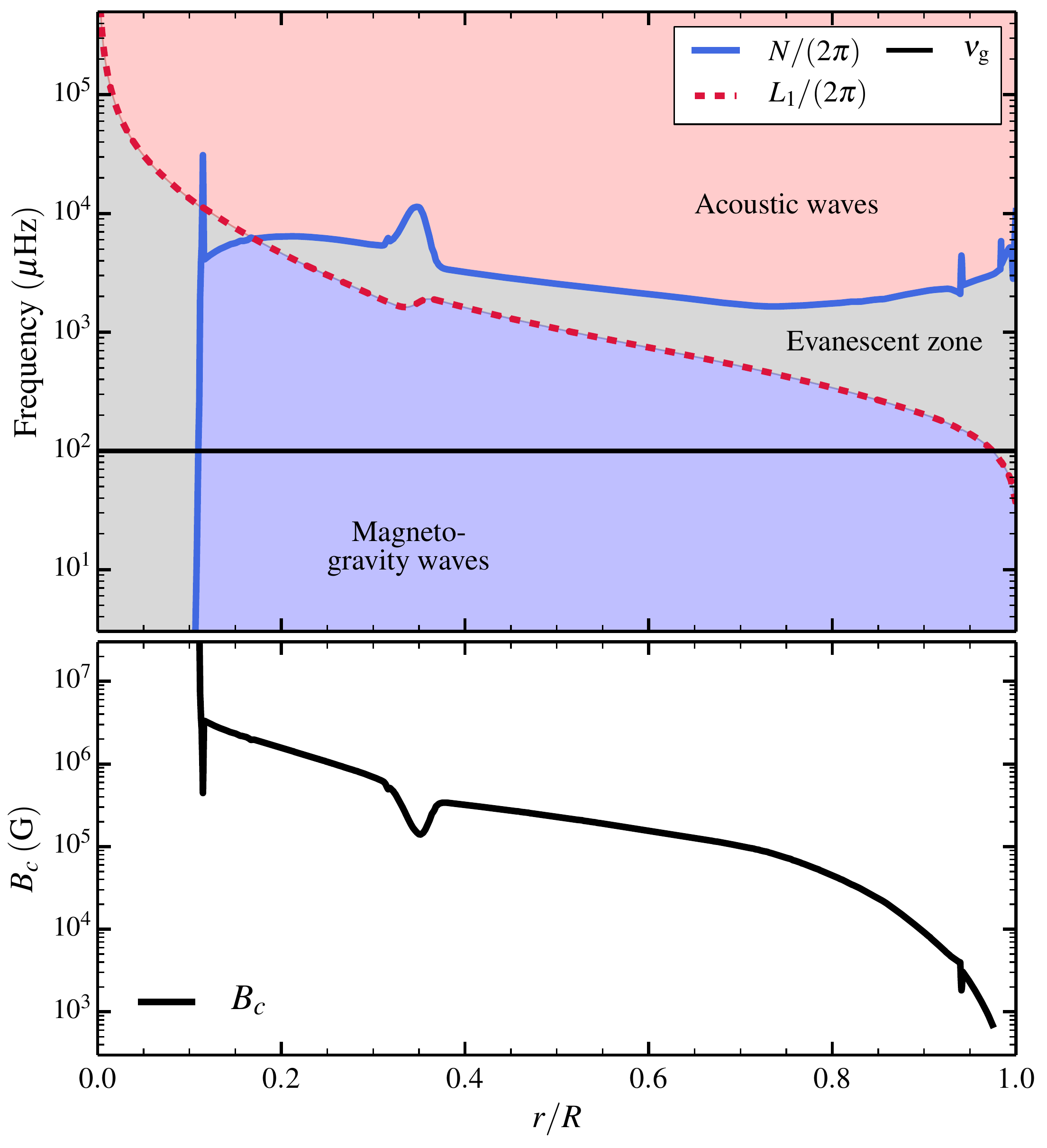}
\caption{\label{fig:sdBProp}
{\bf Top:} Propagation diagram for an sdB model with $M=0.46 \, M_\odot$, $R= 0.22 \, R_\odot$, $T_{\rm eff} = 2.4 \times 10^4 \, {\rm K}$, and $M_{\rm H} = 10^{-3} \, M_\odot$. The horizontal black line indicates the frequency $\nu_{\rm g}$ of a typical sdB g mode excited by the $\kappa$ mechanism. {\bf Bottom:} Critical magnetic field strength $B_c$ (equation \ref{eqn:Bc}) needed to strongly alter dipolar g modes of frequency $\nu_{\rm g}$. We have only plotted $B_c$ in the g mode cavity where equation \ref{eqn:Bc} is valid.%
}
\end{center}
\end{figure}

\begin{figure}[h!]
\begin{center}
\includegraphics[width=0.45\columnwidth]{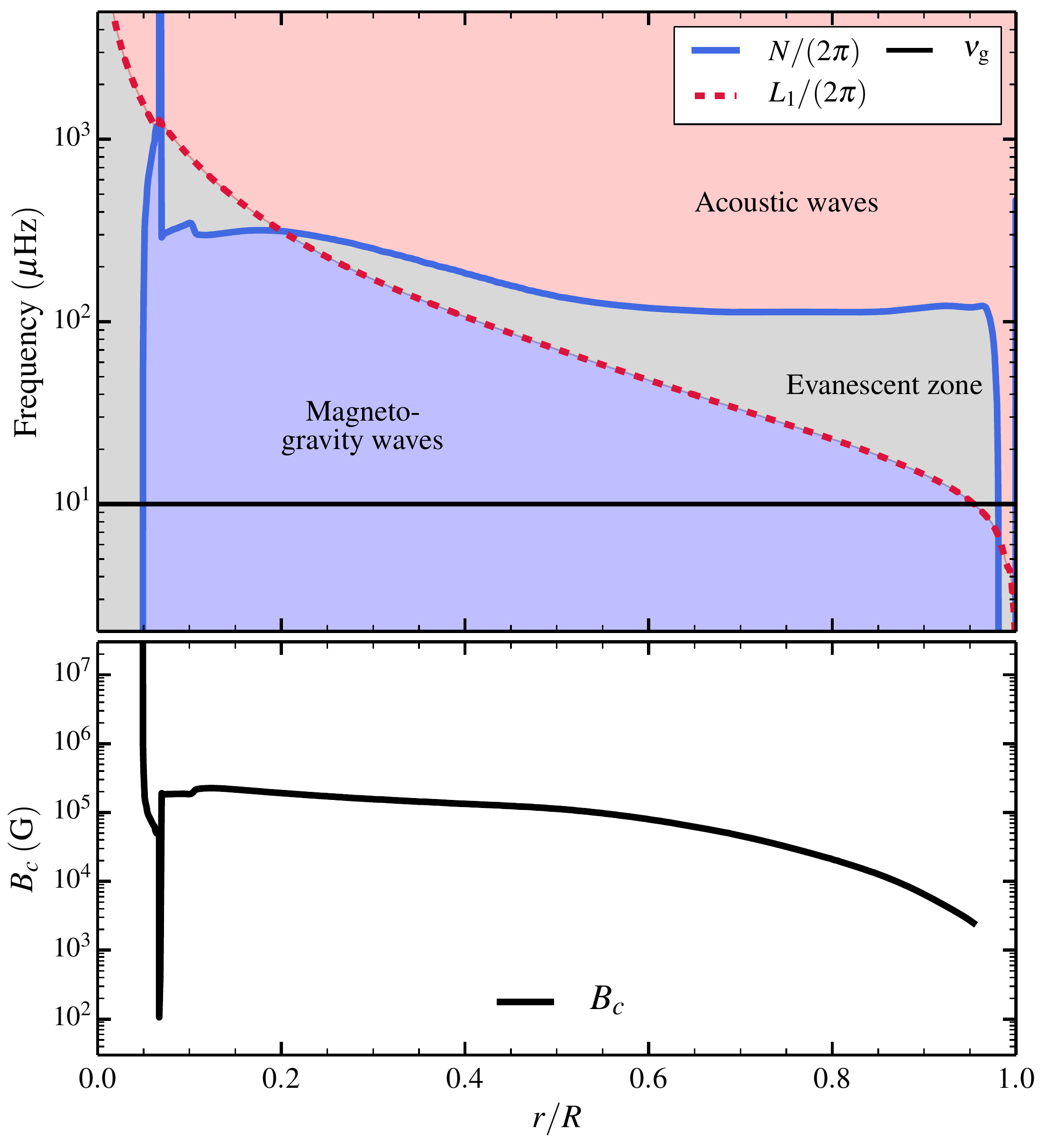}
\caption{\label{fig:GdorProp}
Same as Figure~\ref{fig:sdBProp}, but for a model of a $\gamma$-Dor star. This model has $M=1.6 \, M_\odot$ and $T_{\rm eff} = 6700 \, K$. The horizontal black line indicates the frequency $\nu_{\rm g}$ of a typical $\gamma$-Dor mode. The sharp dip in $B_c$ just outside the convective core is created by the sharp peak in $N^2$ due to composition gradients at the core boundary. This value is sensitive to mixing and overshoot prescriptions and is not reliable.%
}
\end{center}
\end{figure}

\begin{figure}[h!]
\begin{center}
\includegraphics[width=0.45\columnwidth]{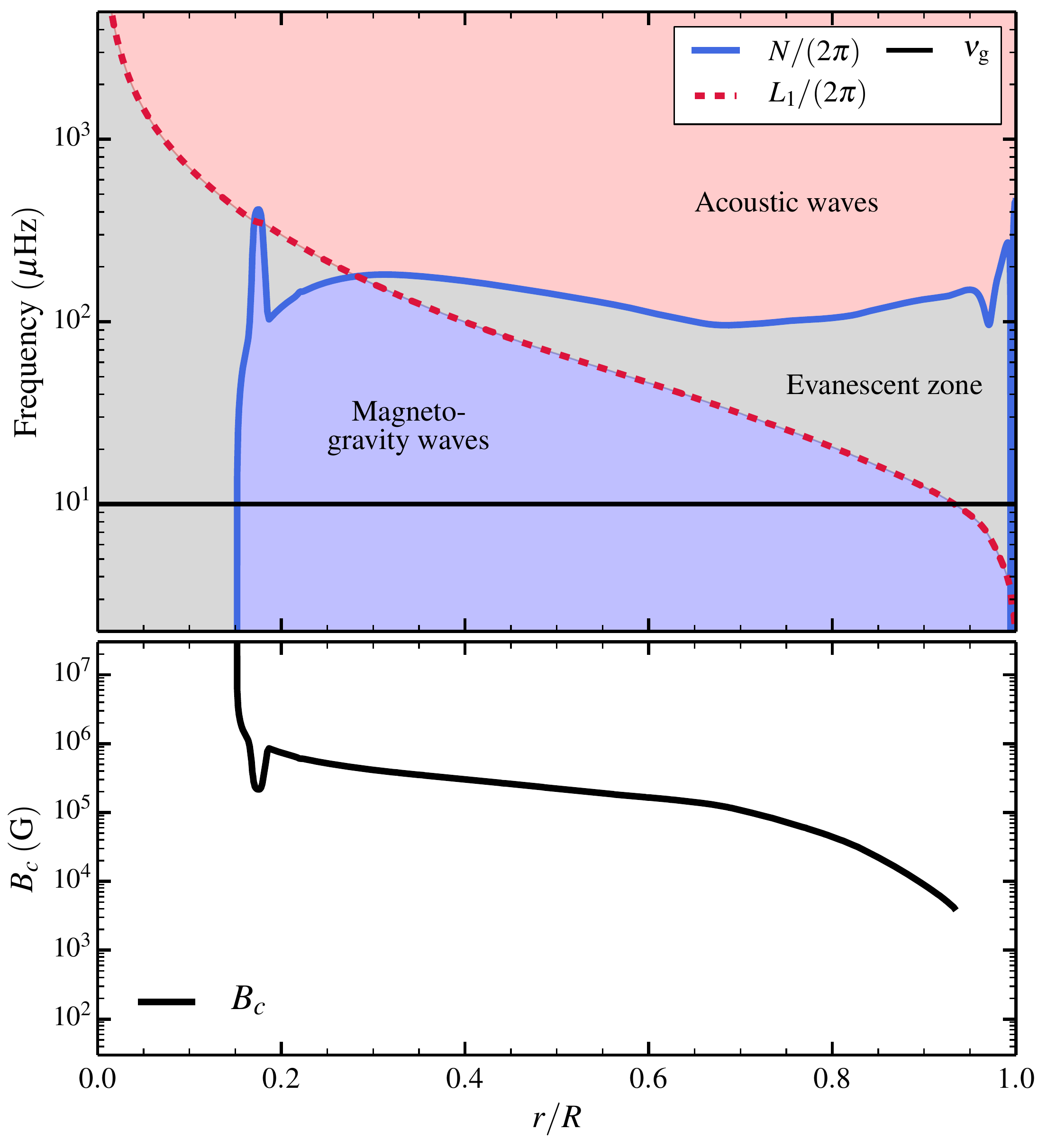}
\caption{\label{fig:SPBProp}
Same as Figure~\ref{fig:sdBProp}, but for a model of a slowly pulsating B star. This model has $M=5 \, M_\odot$ and $T_{\rm eff} = 16000 \, K$. The horizontal black line indicates the frequency $\nu_{\rm g}$ of a typical SPB mode.%
}
\end{center}
\end{figure}

\newpage

\bibliography{bibliography.bib%
}

\end{document}